\documentclass[amsmath,amssymb,reprint,superscriptaddress,floatfix, nofootinbib]{revtex4-2}
\usepackage[dvipsnames]{xcolor}
\usepackage{graphicx}
\usepackage{changepage} 
\usepackage{multirow} 
\usepackage{footmisc}
\usepackage{orcidlink}

\usepackage{hyperref}
\usepackage{float}
\usepackage{soul} 
\usepackage[maxfloats=256]{morefloats}
\usepackage{ulem}
\usepackage{url}
\usepackage{subcaption}
\usepackage{upgreek}

\usepackage[capitalise,noabbrev,nameinlink]{cleveref}
\hypersetup{
citecolor=blue,
 colorlinks=true,
 linkcolor=blue,
 filecolor=magenta,
 urlcolor=blue,
}
\graphicspath{{./figures/}}
\Crefname{equation}{Eq.}{Eqs.}

\usepackage{titlecaps}
\Addlcwords{a and in and of}
\begin{document}

\title{Interacting mesons as degrees of freedom in a chiral model}

\author{Rajesh Kumar\,\orcidlink{0000-0003-2746-3956}}
\email{rkumar6@kent.edu}
\affiliation{Center for Nuclear Research, Department of Physics, Kent State University, Kent, OH 44242 USA}

\author{Joaquin Grefa\,\orcidlink{0000-0001-7590-9364}}
\affiliation{Center for Nuclear Research, Department of Physics, Kent State University, Kent, OH 44242 USA}
\affiliation{Department of Physics, University of Houston, Houston, TX 77204, USA}

\author{Konstantin Maslov}
\affiliation{Cyclotron Institute and Department of Physics and Astronomy, Texas A\&M University, College Station, Texas 77843-3366, USA}
\affiliation{Department of Physics, University of Houston, Houston, TX 77204, USA}

\author{Yuhan Wang}
\affiliation{Center for Nuclear Research, Department of Physics, Kent State University, Kent, OH 44242 USA}

\author{Arvind Kumar}
\affiliation{Department of Physics, Dr. B R Ambedkar National Institute of Technology Jalandhar, Punjab, 144008  INDIA}

\author{Ralf Rapp}
\affiliation{Cyclotron Institute and Department of Physics and Astronomy, Texas A\&M University, College Station, Texas 77843-3366, USA}

\author{Claudia Ratti}
\affiliation{Department of Physics, University of Houston, Houston, TX 77204, USA}

\author{Veronica Dexheimer}
\email{vdexheim@kent.edu}
\affiliation{Center for Nuclear Research, Department of Physics, Kent State University, Kent, OH 44242 USA}

\date{\today}

\begin{abstract}
We study the equation of state of hot and dense hadronic matter using an extended Chiral Mean Field (CMF) model framework where the addition is the inclusion of interactions of thermally excited mesons. This is implemented by calculating the in-medium masses of pseudoscalar and vector mesons, obtained through the explicit chiral symmetry-breaking and vector interaction terms in the Lagrangian, respectively, prior to applying the mean-field approximation. As a result, the in-medium meson contributions generate a feedback term to the CMF's equations of motion, which then modifies the equation of state. 
With this improvement, we quantify the effect on the equation of state of strongly interacting matter through comparisons with state-of-the-art lattice QCD results and other hadronic models like the Hadron Resonance Gas model. We find that the results of the updated hadronic CMF model with an improved meson description (mCMF) provide a better agreement with lattice-QCD data for thermodynamic state variables across a wide range of temperatures and baryon chemical potentials.
\end{abstract}

\maketitle


\section{Introduction}

Studying the properties of hadrons in hot and dense matter is a cornerstone of research in nuclear physics today. In particular, hadronic interactions are an essential ingredient to determining the equation of state (EoS) of hot and dense matter, relevant for heavy-ion collisions and neutron star mergers. Although the hadronic and deconfined quark-gluon phases can be mapped onto the phase diagram of Quantum Chromodynamics (QCD), large portions of it are still largely uncharted \cite{Fukushima:2010bq}. Understanding QCD phase structure is critical not only for understanding the early evolution of the universe, but also for guiding heavy-ion collision (HIC) experiments, and exploring the properties and dynamics of neutron stars and their mergers~\cite{HADES:2019auv,Most:2022wgo}.

While HICs probe the phase structure of excited matter at high temperatures, $T$, and low (net) baryon densities, $n_B$,  compact-star research focuses on the characteristics of extremely dense, cold matter (reaching down to the $T\sim10^{-4}$ MeV scale after a few years of evolution~\cite{Ho:2006uk}). In an ultrarelativistic HIC, a hot fireball forms in the collision zone, where hadronic degrees of freedom dissociate into their fundamental constituents -- quarks and gluons \cite{ALICE:2022wpn}. 
The baryon chemical potential ($\mu_B$) achieved in HICs depends on the energy of the colliding beams. The Beam Energy Scan at the Relativistic Heavy Ion Collider (RHIC) includes fixed-target and collider modes, with energies running from $\sqrt{s_{NN}}=3-200$ GeV 
\cite{Luo:2015doi}, while the HADES experiment at the Helmholtzzentrum fuer Schwerionenforschung (GSI) covers $\sqrt{s_{NN}}=1-3$ GeV \cite{Lorenz:2017hjp}, and FAIR at GSI will have a range of $\sqrt{s_{NN}}=2.9-4.9$ GeV \cite{Durante:2019hzd}; On of the goals of these experiments is to search a the conjectured QCD critical point and corresponding first-order transition line at medium- to low-beam energies \cite{Bzdak:2019pkr}, whereas the experiments at the LHC probe the properties of strongly interacting systems at essentially zero $\mu_B$, similar to the initial stages of the early universe evolution. These experiments yield detailed measurements of particle production, enabling the extraction of the thermodynamic and collective properties of the systems created in HICs.

Despite decades of intense research, the phase structure of strongly interacting matter remains elusive to a large extent. Exceptions include the regime near cold, saturated nuclear matter and the smooth crossover transition from hadronic to partonic degrees of freedom at zero $\mu_B$~\cite{Aoki:2006we}. In the latter regime, the restoration of chiral symmetry signaled by the peak of chiral susceptibility with a pseudocritical temperature of approximately $T_{\rm pc}\approx158$ MeV \cite{Borsanyi:2020fev} is fundamental in the description of the broad crossover at vanishing density \cite{Borsanyi:2010bp}. At low $T$ and high $n_{B}$, a transition to cold, deconfined quark matter state is expected, with a possible onset of new condensation phenomena related to color-superconductivity, including a first-order phase transition~\cite{Rapp:1999qa,Alford:2007xm}. This transition is part of the broader phenomena of the emergence of a quark-gluon plasma (QGP) phase at high $T$ and $\mu_B$~\cite{Shuryak:1980tp}. For a comprehensive review of theoretical and experimental constraints for the equation of state of dense and hot matter, we refer the reader to Ref.~\cite{MUSES:2023hyz}.

Although QCD is the first-principle theory for strongly interacting matter, the strong coupling nature of QCD in the phase transition regions at the pertinent $T$ and $\mu_{B}$ values in heavy-ion collisions, neutron stars, and their mergers renders perturbative calculations not applicable. At vanishing $n_B$, QCD thermodynamics can be simulated numerically on the lattice; however, such calculations are hindered by the sign problem at finite $\mu_B$ \cite{Ratti:2018ksb}. One can bypass this obstacle by reconstructing the EoS using susceptibilities, computed on the lattice at $\mu_{B}=0$, through a Taylor expansion \cite{Ratti:2022qgf}, and most recently through an alternative summation scheme where the EoS is extrapolated to finite $n_B$  through a $\mu_B$ dependent $T$-expansion~\cite{Borsanyi:2021sxv}. Unfortunately, these approaches only cover a small region in the phase diagram that restricts thermodynamic calculations to $\mu_{B}/T\leq 3.5$.

The absence of a first-principle EoS for strongly interacting matter at intermediate-to-high $\mu_B$ and low $T$ remains a critical challenge, especially for neutron star and neutron star merger studies. Consequently, effective field-theoretical approaches to describe strongly interacting matter that respect QCD symmetries and reasonably capture the known phenomenology of strong interactions are essential for exploring the various phases of QCD. While relativistic mean-field models with baryon and quark degrees of freedom are widely used to compute the neutron-star equation of state (including mesons only to mediate interactions), thermal models with mesons, such as the Hadron Resonance Gas (HRG) model, are extensively used by the HIC community. In this framework, the system is described by a non-interacting gas of hadrons where the thermal excitation of resonances accounts for the interactions present in the system; we will refer to this as the ideal HRG model (HRG$_{ideal}$). {However, it is important to capture the short-range repulsion and long- and medium-range attraction of the nucleon-nucleon interaction and reproduce the nuclear liquid-gas phase transition. One way to do that is to employ the HRG model with van-der-Waals interactions (HRG$_{\rm VDW}$), which reproduces the saturation properties of nuclear matter~\cite{Vovchenko:2016rkn}.  Another possibility is to use relativistic mean-field models, which can be extended to higher hadronic multiplets to adequately compare with the lattice QCD data~\cite{Khvorostukhin:2006ih, Khvorostukhin:2008xn, Steinert:2018zni}.

In chiral effective models, baryon masses are generated through their interactions with the medium. Small explicit chiral-symmetry breaking effects are incorporated to ensure that pseudoscalar mesons acquire small but finite masses. The Chiral Mean Field (CMF) model, based on a nonlinear realization of the SU(3) sigma model, accounts for hadronic (and quark) interactions mediated by meson mean fields and can be constrained using lattice-QCD results, nuclear experiments, and astrophysical observations \cite{Dexheimer:2008ax,Dexheimer:2009hi,Kumar:2024owe, Cruz-Camacho:2024odu}. There is also an alternative CMF model that includes the conjectured chiral partners of the baryon octet and non-interacting contributions from all established hadronic resonances~\cite{Steinheimer:2011ea,Motornenko:2019arp}. Other examples of effective chiral models include the Polyakov Nambu-Jona-Lasinio (PNJL) model which is useful to study quark matter and the QGP/hadronic phase transition \cite{Fukushima:2003fw,Ratti:2005jh}, and the linear sigma model \cite{Lenaghan:2000ey} with its Polyakov extension \cite{Tawfik:2014gga} that  has been employed to investigate hadron and quark dynamics. However, one of the advantages of the nonlinear realization of the SU(3) sigma model, that CMF is based on, is the inclusion of separate explicit chiral symmetry-breaking terms that preserve partially conserved axial currents (PCAC) and decouple strange from non-strange condensates. This occurs when the pseudoscalar mesons become the parameters of the chiral transformation and, as a result, only appear if the symmetry is explicitly broken or in terms with derivatives of the fields \cite{Papazoglou:1998vr}.

In the CMF model, the mean-field approximation is introduced
to simplify the full quantum-operator fields from the nonlinear realization of the SU(3) sigma model in order to obtain a Lagrangian that can be solved straightforwardly using classical methods~\cite{Reinhard:1989zi,Serot:1984ey} where mesonic quantum field operators are replaced with classical expectation values, assuming time-independent, homogeneous, and isotropic infinite matter. Additionally, because of parity conservation, pseudoscalar meson mean-field expectation values vanish, e.g., $\left<\pi\right>=0$. Additionally, the no-sea approximation is employed, which assumes that only valence (positive-energy) states contribute, ignoring negative-energy solutions from the Dirac sea. This eliminates vacuum polarization effects, simplifying calculations while maintaining the essential physics of nuclear interactions. Furthermore, a deconfinement mechanism is included via  a Polyakov-loop-like field and potential, which allows a change of degrees of freedom from hadrons to quarks \cite{Dexheimer:2009hi}.

Until now, the CMF model only included baryons and/or quarks as possible degrees of freedom interacting through meson mean-fields. The model was also supplemented by the contribution of a non-interacting ideal gas of pseudoscalar and vector mesons with their vacuum masses, which gives an important contribution at large temperatures.
The latter were referred to as thermal mesons. The model~\cite{Dexheimer:2009hi} included the full temperature and chemical potential dependence of meson mean fields and their effect on the baryon and quark masses through scalar mesons ($\sigma, \delta, \zeta$), but not on the thermal meson masses. Furthermore, it was assumed that the vector mesons ($\omega, \rho, \phi$) have the same mass. This mass degeneracy was later broken in the CMF model by adding a chirally invariant term to the vector meson Lagrangian, resulting in a redefinition of the vector meson fields~\cite{Kumar:2024owe} but still without taking into account the 
thermal-meson mass dependence on the in-medium mean fields. This assumption limited the model’s ability to capture the EoS at large $T$ and small $\mu_{B}$ because under these conditions the mesons are expected to be the dominant degrees of freedom.

In-medium properties of mesons have been investigated, in particular, due to their relevance for neutron star phenomenology \cite{Kaplan:1986yq}, as well as for relativistic HICs \cite{Yan:2024gwp}. In the CMF framework the meson mean fields are the force carriers, and within chiral approaches, the particle masses (now not only for baryons but also for mesons) are dynamically generated from the interaction with the medium and depend on $T$ and $n_B$ \cite{Rapp:2009yu,RajeBhageerathi:2018wap,Zschiesche:2002zr}.
While there are studies related to the in-medium properties of mesons using the non-linear chiral SU(3) model utilizing different parametrizations, the focus has been on understanding how meson properties, such as mass and decay width, evolve in dense and hot nuclear matter~\cite{Kumar:2018ujk,Kumar:2020vys, Mishra:2004te}. In these studies, a hybrid approach was taken by either calculating pseudoscalar meson properties such as in-medium mass by taking baryon densities and mean meson fields (from the calculations without in-medium meson masses) as an input in their dispersion relations~\cite{Kumar:2020vys,Mishra:2004te}, or combining the model with QCD sum rules to compute meson properties~\cite{Kumar:2018ujk}. In these works, the meson masses were modified due to contributions from the terms with derivative couplings (such as the Weinberg-Tomozawa term, pseudoscalar-scalar meson interactions, and pseudoscalar-baryon interactions) and explicit symmetry-breaking terms. 

Unlike these previous studies, the work in this paper extends the CMF model and focuses solely on the explicit symmetry-breaking and vector interaction terms, excluding derivative couplings and self-interactions of pseudoscalar mesons (which is more in line with the mean-field approximation). This approach, however, does not account for any the meson decay or collisional widths, and is thus limited to the quasiparticle approximation.
In this work, we examine in-medium meson masses before applying the mean-field approximation in the CMF framework, for the first time. As a consequence, the field-dependent in-medium contributions from the thermal mesons introduce a feedback term in the mean-field equations of motion. We also describe in detail how the interacting thermal mesons are included within the CMF formalism for the first time. We present our results utilizing the new hadronic CMF model with improved {\bf{meson}} description (mCMF) for the EoS of isospin-symmetric matter at finite $T$ and $\mu_B$ without constraints on the net strangeness, i.e., for the electric charge and strangeness chemical potentials $\mu_{Q}= \mu_{S}=0$. In addition to the pseudoscalar and vector mesons, our degrees of freedom include the baryon octet and decuplet.

This paper is organized as follows. In Section~\ref{sec:formalism}, we present the basic expressions of the mCMF formalism. In Section~\ref{sec:results}, we show our results regarding the contribution of the interacting thermal mesons to the mean fields, to their in-medium mass and number densities. Additionally, we present thermodynamic quantities for hot and dense hadronic matter resulting from the mCMF model and how they compare to the state-of-the-art lattice QCD data and HRG models (including partial pressures). In Section~\ref{sec:conclusions}, we conclude by summarizing our results and giving a brief outline of future perspectives based on the results presented in this work.
 
\section{Formalism}
\label{sec:formalism}
The following Section briefly presents the formalism for the mCMF model. We show new expressions for the equations of motion, relevant thermodynamical quantities, and in-medium masses of pseudoscalar and vector mesons.
\subsection{Hadronic CMF model with improved meson description (mCMF model)}\label{sec:cmf_model}  

The Chiral Mean Field (CMF) model is a relativistic framework developed to describe the thermodynamic properties and particle populations of hadronic and quark matter in dense and hot environments~\cite{Dexheimer:2009hi}. The CMF incorporates essential QCD properties such as broken scale invariance and chiral symmetry breaking~\cite{Weinberg:1968de, Coleman:1969sm}. It employs a non-linear realization of the SU(3) chiral symmetry with baryons coupled to  exchange mesons via a Yukawa-type coupling to describe the strong force. Key mesons include the scalar fields $\sigma$, $\zeta$, and $\delta$, which govern attractive forces, and the vector fields $\omega$, $\phi$, and $\rho$, which handle short-range repulsive interactions. In the mean-field approximation (MFA), we assume infinite baryonic matter that is homogeneous and isotropic, characterized by definite positive parity and zero charge. Therefore, only meson mean fields with positive parity, specifically, scalar mesons and the time-like component of vector mesons are nonzero. Additionally, only those with a zero third isospin component, corresponding to the mesons along the diagonal of the matrices $X$ (see \Cref{eq:X_matrix}) and $V_\mu$ (see \Cref{eq:V_matrix}), remain non-vanishing. The meson mean fields with negative parity, including the space-like component of vector mesons, the time-like component of axial-vector mesons, and pseudoscalars, do not follow parity conservation and, therefore, there is no source term for these mesons in the mean-field infinite baryonic matter. Moreover, within MFA approximation, fluctuations around the constant ground-state expectation values of the scalar and vector field operators are disregarded and, as a consequence, all scalar ($\sigma$, $\zeta$, and $\delta$) and vector ($\omega_0$, $\phi_0$, and $\rho_0$) mesons become time and coordinate-independent fields~\cite{Cruz-Camacho:2024odu}. 

\begin{table}[t!]
\centering
\caption{The values of the hyperon and Delta-baryon potentials reproduced in the current work.}
\begin{tabular}{c c}
\hline \hline
\text{Baryon} & $U_B$ (MeV) \\
\hline \hline
$\Lambda$ & -28.06 \\
$\Sigma$ & 5.51 \\
$\Xi$ & -15.67 \\
$\Delta$ & -82 \\
$\Sigma^*$ & 5.70 \\
$\Xi^*$ & -8.32 \\
$\Omega^-$ & 28.70 \\
\hline \hline
\end{tabular}
\label{tab:value_param_constraints_hyperon} 
\end{table}

The hadronic CMF parameters were fitted to reproduce standard nuclear properties at saturation for isospin-symmetric matter, the symmetry energy also at saturation, baryon and meson vacuum masses, in addition to decay constants and other physical results at $T=0$. See Tab.~XII in Ref.~\cite{Cruz-Camacho:2024odu} for details and Ref.~\cite{Kumar:2024owe} for specific values for the field-redefined C4 vector coupling.
Furthermore, the parameters relevant to hyperon and Delta-baryon interactions in the medium can be constrained by the available hypernuclear data and by, e.g., the data on photoabsorption of nuclei~\cite{MUSES:2023hyz}. These constraints are formulated in terms of the baryon potentials, $U_i=m_i^*-m_i+g_{\omega i} \omega+g_{\rho i} \rho+g_{\phi i} \phi$ in nuclear matter calculated at saturation density for isospin-symmetric matter at $T=0$. Within the field-redefined C4 vector coupling parametrization, the parameters $m^{HO}_3=0.8061$  (see \Cref{eq:mdh}) are chosen to reproduce the hyperon potentials of the strange baryon octet, whereas the hyperon potentials of the baryon decuplet are reproduced by the following parameters, $g^X_D$=3.54, $\alpha_{DX}=0.248$, $V_\Delta=1$, $m_0=190$ MeV, and  $m^{HD}_3=1.9$ (see again Ref.~\cite{Cruz-Camacho:2024odu} for more details about these parameters). The corresponding hyperon and Delta-baryon potentials are tabulated in \Cref{tab:value_param_constraints_hyperon}.  The remaining parameters associated with the scalar and vector sectors of the CMF model can be found in Ref.~\cite{Cruz-Camacho:2024odu}. 

The CMF model is consistent with low energy nuclear  physics experimental data at nuclear saturation density~\cite{Dexheimer:2008ax} and provides a relativistic framework for studying the properties of strongly interacting matter in both heavy-ion collisions~\cite{Steinheimer:2007iy,Steinheimer:2009nn} and compact stars~\cite{Dexheimer:2008ax,Dexheimer:2015qha,Dexheimer:2018dhb}. It can also describe quark matter by including quark deconfinement through an order parameter,  $\Phi$, which accompanies a Polyakov loop-like potential~\cite{Dexheimer:2009hi}. The model can accommodate various conditions, including different $T$, $\mu_B$, and magnetic fields~\cite{Dexheimer:2011pz,Franzon:2015sya,Dexheimer:2021sxs,Marquez:2022fzh,Peterson:2023bmr}, making it a valuable tool for exploring different regions of the QCD phase diagram~\cite{Dexheimer:2009hi,Hempel:2013tfa,Roark:2018uls,Aryal:2020ocm,Peterson:2023bmr,Kumar:2024owe}. In Ref.~\cite{Cruz-Camacho:2024odu}, we provide a comprehensive review of the CMF model, focusing at $T=0$, while deriving all relevant expressions and discussing its key features and relevance for understanding dense nuclear matter (independently) for different conserved charges: baryon number $B$, electric charge $Q$, and strangeness $S$.
At finite $T$, in Ref.~\cite{Kumar:2024owe}, we focus on a new fit of the field-redefined CMF model to better describe the most recent lattice QCD results. 

The hadronic CMF model Lagrangian takes the form~\cite{Cruz-Camacho:2024odu}
\begin{equation}
\mathcal{L}_{\rm CMF} = \mathcal{L}_{\rm kin} + \mathcal{L}_{\rm int} + \mathcal{L}_{\rm scal} + \mathcal{L}_{\rm vec} + \mathcal{L}_{m_0} + \mathcal{L}_{\rm esb}-  \mathcal{L}^{\rm M. F.}_{\rm vacuum}\,.
\end{equation}
Here, \( \mathcal{L}_{\rm kin} \) represents the kinetic terms for the baryon octet and decuplet, \( \mathcal{L}_{\rm int} \) accounts for interactions between these fermions and the $6$ scalar and vector meson mean fields mentioned above. The self-interactions of scalar mesons are included in \( \mathcal{L}_{\rm scal} \), while \( \mathcal{L}_{\rm vec} \) contains terms that determine vector meson masses and self-interactions up to quartic order. The $\mathcal{L}_{m_0}$ term is added in order to fit the compressibility for the C4 vector parametrization \cite{Cruz-Camacho:2024odu}, which we use in this work. The explicit chiral symmetry breaking term \( \mathcal{L}_{\rm esb} \) (discussed in the following) enables the model to generate the masses of Goldstone bosons and also to reproduce observed hyperon potentials (see Eq. (18) of Ref.~\cite{Cruz-Camacho:2024odu}  for the detailed discussion). Finally, $\mathcal{L}^{\rm M. F.}_{\rm {vacuum}}$ is the constant vacuum term containing scalar  meson mean-field self-interactions. 

In this study, we improve the field-redefined CMF model of Ref.~\cite{Kumar:2024owe} by including a thermal meson contribution that includes interactions, meaning that the pseudoscalar and vector meson masses are now dependent on the in-medium mean fields. As a first step toward a more realistic EoS, in this work, we consider these improvements in the hadronic version of the CMF model, where we neglect all quark contributions to the Lagrangian. The resulting hadronic (\textit{H}) grand canonical potential density can be written as:
\begin{align}
\frac{{\Omega}^{H}}{V}&=U+\frac{{\Omega}^B_{\rm th}}{V}+\frac{{\Omega}^M_{\rm th}}{V}\,,
\label{eq:hadronic_thermo_pot}
\end{align}
where $V$ stands for volume and $U$ represents the interaction potential energy density, which in the MFA can be expressed in terms of Lagrangian 
\begin{equation}
U=-\mathcal{L}_{\rm kin} - \mathcal{L}_{\rm int} - \mathcal{L}_{\rm scal} - \mathcal{L}_{\rm vec}-\mathcal{L}_{m_0} - \mathcal{L}_{\rm esb} +  \mathcal{L}^{\rm M. F.}_{\rm vacuum}\,,
\label{eq:U_meson_pot}
\end{equation}
and the thermal contributions from baryons (and antibaryons) can be expressed as
\begin{align}
\frac{{\Omega}^B_{\rm th}}{V}&=-T \sum_{i \in { B}} \frac{\gamma_i}{2 \pi^2} \int dk k^2\left(\ln \left[1+e^{-\frac{1}{T}\left(E_i^*(k)-\mu_i^*\right)}\right]\right.\nonumber \\
&\left. +\ln \left[1+e^{-\frac{1}{T}\left(E_i^*(k)+\mu_i^*\right)}\right]\right) \,,
\label{eq:baryonic_thermo_pot}
\end{align}
where $i$=$p,n, \Lambda, \Sigma, \Xi,  \Delta, \Sigma^*, \Xi^*, \Omega^-$, and for mesons
\begin{align}
\frac{{\Omega}^M_{\rm th}}{V}&=T \sum_{i \in { M}} \frac{\gamma_i}{2 \pi^2} \int dk k^2 \ln \left[1-e^{-\frac{1}{T}\left(E_i^*(k)-\mu^*_i\right)}\right]\,,
\label{eq:mesonic_thermo_pot}
\end{align}
where $i$=$\pi, \eta, \eta^{\prime}, K, {\omega}, \rho, K^*, {\phi}$~\footnote{\label{note1}We are treating each meson and anti-meson as different particles - unlike baryons that are treated as the same particle with a different distribution function.}. Also, $\gamma_i$ is the spin degeneracy, $k$ is the momentum, $E_i^*=\sqrt{k^2+m^{*^2}_i}$ is the effective quasiparticle dispersion relation, and $m^*_i$ denotes the in-medium (or effective) mass of particle $i$, baryon or meson. The effect of vector mean fields on baryon dispersion relations is captured by the effective chemical potentials
\begin{gather}
    \mu_i^*=\mu_i-g_{\omega i}\omega_0-g_{\rho i}\rho_0-g_{\phi i}\phi_0,\,
\end{gather}
where $g_{\omega i}$, $g_{\rho i}$, and $g_{\phi i}$ are the coupling constants of a baryon $i$ with the corresponding mean-field. The particle chemical potential is given by $\mu_i=B_i\mu_B+S_i\mu_S+Q_i\mu_Q\,,$ where $B_i$, $S_i$ and $Q_i$ (antibaryons are included with the same $\mu$ as baryons, but summed separately following a different distribution function) denote the particle's baryon, strange and electric charge number, respectively. Note that for mesons, $\mu_i^*=\mu_i$ since no interactions of mesons with vector mean fields are present in our approach. 

For baryons, the in-medium masses can be obtained from the following expression
\begin{equation}
m_i^*=g_{\sigma i}\sigma+g_{\zeta i}\zeta+g_{\delta i}\delta+\Delta m_i\,,
\label{eq:emh}
\end{equation}
where $g$ represents the coupling constants of baryons with the scalar mean fields. The parameter $\Delta m_i$ accounts for contributions from several sources: the bare mass $m_0$ associated with the baryon octet and decuplet, the explicit symmetry-breaking parameter $m^{HO}_3$ for hyperons ($HO$) from the octet, and the explicit symmetry-breaking parameter $m^{HD}_3$ for hyperons ($HD$) from the decuplet:
\begin{align}
&\Delta m_N =m_0\,, \nonumber \\
&\Delta m_\Lambda = \Delta m_\Sigma = \Delta m_\Xi = m_0  - m^{HO}_3 \left( \sqrt{2} \sigma_0 + \zeta_0  \right)\,,  \nonumber\\ 
&\Delta m_{\Delta}= m_0\,, \label{eq:mdh} \\ 
&\Delta m_{\Sigma^*} =   \Delta m_{\Xi^*}= m_0 -m^{HD}_{3}(\sqrt{2} \sigma_0+\zeta_0)\,, \nonumber\\ 
&\Delta m_\Omega = m_0-\frac{3}{2}m^{HD}_{3}(\sqrt{2} \sigma_0+\zeta_0)\,, \nonumber
\end{align}
where $\sigma_0 (=-f_\pi)$ and $\zeta_0 (=\frac{f_\pi}{\sqrt{2}}-\sqrt{2} f_K)$ are the vacuum expectation values of the scalar mean fields fitted to experimental values of decay constants of pions and kaons. The vacuum expectation value of the scalar-isovector $\delta$ mean-field is zero.

The inclusion of thermal mesons with field-dependent masses influences the mean-field equations of motion, as they introduce back-reaction (or feedback) terms. The equations of motion are then adjusted accordingly:
\begin{align}
\frac{\partial\left({\Omega}^H / V\right)}{\partial \sigma} &= {\frac{\partial{U}}{\partial \sigma}} + {\frac{{\Omega}^B_{\rm th}/V}{\partial \sigma}}\nonumber\\&+
\sum_{i \in { M}} \frac{\gamma_i}{2\pi^2}\int dk k^2 \frac{m_{i}^{*}}{E_{i}^{*}} \frac{\partial m_{i}^{*}}{\partial \sigma} f^M_{+_i} \,,\label{eq:eom_sig} \\
\frac{\partial\left({\Omega}^H / V\right)}{\partial \zeta} &= {\frac{\partial{U}}{\partial \zeta}} + {\frac{{\Omega}^B_{\rm th}/V}{\partial \zeta}}\nonumber\\&+\sum_{i \in { M}} \frac{\gamma_i}{2\pi^2} \int dk k^2 \frac{m_{i}^{*}}{E_{i}^{*}} \frac{\partial m_{i}^{*}}{\partial \zeta} f^M_{+_i} \,,\label{eq:eom_zet} \\
\frac{\partial\left({\Omega}^H / V\right)}{\partial \delta} &= {\frac{\partial{U}}{\partial \delta}} + {\frac{{\Omega}^B_{\rm th}/V}{\partial \delta}}\nonumber\\&+\sum_{i \in { M}} \frac{\gamma_i}{2\pi^2} \int dk k^2\frac{m_{i}^{*}}{E_{i}^{*}} \frac{\partial m_{i}^{*}}{\partial \delta} f^M_{+_i} \,,\label{eq:eom_del} \\
\frac{\partial\left({\Omega}^H / V\right)}{\partial \omega_0} &= {\frac{\partial{U}}{\partial \omega_0}} + {\frac{{\Omega}^B_{\rm th}/V}{\partial \omega_0}}\nonumber\\&+
\sum_{i \in { M}}\frac{\gamma_i}{2\pi^2} \int dk k^2 \frac{m_{i}^{*}}{E_{i}^{*}} \frac{\partial m_{i}^{*}}{\partial {\omega_0}} f^M_{+_i} \,,\label{eq:eom_ome}
\end{align}
\begin{align}
\frac{\partial\left({\Omega}^H / V\right)}{\partial \rho_0} &= {\frac{\partial{U}}{\partial \rho_0}} + {\frac{{\Omega}^B_{\rm th}/V}{\partial \rho_0}}\nonumber\\&+\sum_{i \in { M}}\frac{\gamma_i}{2\pi^2}\int dk k^2\frac{m_{i}^{*}}{E_{i}^{*}} \frac{\partial m_{i}^{*}}{\partial \rho_0} f^M_{+_i}\,, \label{eq:eom_rho} \\
\frac{\partial\left({\Omega}^H / V\right)}{\partial \phi_0} &= {\frac{\partial{U}}{\partial \phi_0}} + {\frac{{\Omega}^B_{\rm th}/V}{\partial \phi_0}}\nonumber\\&+
\sum_{i \in { M}}\frac{\gamma_i}{2\pi^2} \int dk k^2\frac{m_{i}^{*}}{E_{i}^{*}} \frac{\partial m_{i}^{*}}{\partial {\phi_0}} f^M_{+_i}\,. \label{eq:eom_phi}
\end{align}
We defer the discussion of the explicit expressions for the field-dependent in-medium masses of mesons to the next subsection. In general form, by rearranging the above equations using the expression for the scalar density of mesons (defined in the following in \cref{eq:meson_observables}) and writing it compactly, we get
\begin{equation}
\frac{\partial\left({\Omega}^H / V\right)}{\partial {\vartheta}} = \frac{\partial({\Omega^{orig}} / V)}{\partial {\vartheta}} + \sum_{i \in { M}} n^M_{s_i}\frac{\partial m_{i}^{*}}{\partial {\vartheta}}\,,
\label{eq:compact_EoM}
\end{equation}
where $\vartheta=\sigma,\,\zeta,\,\delta,\,\omega_{0},\,\rho_0,\,\phi_{0}$. From now on, for simplicity, we drop the subscript 0 from the vector meson fields. In the above equation, the first term on the right-hand side represents the original CMF model's equations of motion (see Eq.~46 in Ref.~\cite{Cruz-Camacho:2024odu}) with thermal contribution from interacting baryons, and the second term represents the thermal contributions from interacting mesons introduced in this work. Note that thermal contributions from non-interacting mesons would not contribute to \Cref{eq:eom_sig,eq:eom_zet,eq:eom_del,eq:eom_ome,eq:eom_rho,eq:eom_phi,eq:compact_EoM}, as their masses are independent of the mean fields.

The baryonic number density, scalar density, energy density, pressure, and entropy density are defined respectively (as usual) as
\begin{align}
n^B_i&=\frac{\gamma_i}{2\pi^2}\int_{0}^{\infty}dk k^2 \left(f^B_{+_i}-f^B_{-_i}\right)\,,\nonumber  \\
n^B_{s_i}&=\frac{\gamma_i}{2\pi^2}\int_{0}^{\infty}dk k^2 \frac{m^*_i}{E^*_i} \left(f^B_{+_i}+f^B_{-_i}\right)\,, \nonumber \\
\varepsilon^B_i&=\frac{\gamma_i}{2\pi^2}\int_{0}^{\infty}dk k^2 E^*_i \left(f^B_{+_i}+f^B_{-_i}\right)\,,\nonumber \nonumber \\
P^B_i&=\frac{1}{3}\frac{\gamma_i}{2\pi^2}\int_{0}^{\infty}dk \frac{k^4} {E^*_i} \left(f^B_{+_i}+f^B_{-_i}\right)\,,\nonumber \\
s^B_i&=\dfrac{\gamma_i}{2\pi^2}\int_0^\infty dk k^2\Bigg[f^B_{+_i}\ln\left(\dfrac{1}{f^B_{+_i}}\right)
+f^B_{-_i}\ln\left(\dfrac{1}{f^B_{-_i}}\right)\nonumber\\
+(1&-f^B_{+_i})\ln\left(\dfrac{1}{1-f^B_{+_i}}\right)+(1-f^B_{-_i})\ln\left(\dfrac{1}{1-f^B_{-_i}}\right)\Bigg]\,.
\label{eq:baryon_observables}
\end{align}
Here, $f^B_{\pm_i}$ represents the Fermi-Dirac distribution function given by
$
f^B_{\pm_i}=\left[{e^{(E_i^*\mp\mu_i^*)/T}+1}\right]^{-1}
$.
where the $\pm$ stands for particle vs. antiparticle.
On the other hand, the mesonic number density, scalar density, energy density, pressure, and entropy density are defined (as usual) respectively as
\begin{align}
n^M_i&=\frac{\gamma_i}{2\pi^2}\int_{0}^{\infty}dk k^2 f^M_{+_i}\,, \nonumber \\
n^M_{s_i}&=\frac{\gamma_i}{2\pi^2}\int_{0}^{\infty}dk k^2 \frac{m^*_i}{E^*_i} f^M_{+_i}\,,\nonumber \\
\varepsilon^M_i&=\frac{\gamma_i}{2\pi^2}\int_{0}^{\infty}dk k^2 E^*_i f^M_{+_i}\,,\nonumber \\
P^M_i&=\frac{1}{3}\frac{\gamma_i}{2\pi^2}\int_{0}^{\infty}dk \frac{k^4} {E^*_i} f^M_{+_i}\,,\nonumber \\
s^M_i&=\frac{\gamma_i}{2\pi^2}\int_{0}^{\infty}dk k^2 \bigg[(f^M_{+_i}+1)\ln(f^M_{+_i}+1)-f^M_{+_i} \ln f^M_{+_i}\bigg]\,,
\label{eq:meson_observables}
\end{align}
where,  $f^M_{+_i}$ represents the Bose-Einstein distribution function given by
$
f^M_{+_i}=\left[{e^{(E_i^*-\mu_i^*)/T}-1}\right]^{-1}
\label{eq:BE_distribution_function}
$
with $``+"$ standing for particle, where for antiparticles $\mu^*_i(=\mu_i)$ is negative$^{\ref{note1}}$. Note that for a non-interacting gas of thermal mesons  $n^M_{s_i}=0$.

The total energy density, pressure, and entropy density of the hadrons contain several different contributions. First the kinetic contributions of the baryons, then the kinetic contribution of the mesons having medium dependent mass,  vector mean-field contribution to the energy density only~\cite{Cruz-Camacho:2024odu}, scalar mean-field contribution to the energy density and pressure only, and finally the  vacuum correction to the energy density and pressure only  
\begin{align}
\varepsilon_{\rm total}&=\sum_{i \in { B}} \varepsilon^B_i+\sum_{i \in { M}} \varepsilon^M_i+\varepsilon_{\rm int}-\left(\mathcal{L}_{\rm mesons}^{\rm M.F.}-\mathcal{L}_{\rm vacuum}^{\rm M.F.}\right)\,, \nonumber \\
P_{\rm total}&=\sum_{i \in { B}} P^B_i+\sum_{i \in { M}} P^M_i
+(\mathcal{L}_{\rm mesons}^{\rm M.F.}-\mathcal{L}_{\rm vacuum}^{\rm M.F.})\,,\nonumber \\
s_{\rm total}&=\sum_{i \in { B}} s^B_i+\sum_{i \in { M}} s^M_i\,.
\label{eq:hadron_observables_mod}
\end{align}
See Eqs. 58 and 62 in Ref.~\cite{Cruz-Camacho:2024odu} for more details about the vector mean-field and vacuum terms, respectively. 

\subsection{In-medium masses of mesons}
\label{sec:thermal_meson_mass}

The second derivative of the interaction potential $U$ (\Cref{eq:U_meson_pot}) gives information regarding the curvature of $U$ with respect to field fluctuations. When evaluated at the vacuum value, this curvature determines the effective masses of small fluctuations around the vacuum.
This is the mechanism we use to calculate the in-medium masses of thermal vector and pseudoscalar mesons $m^{*}_{\varphi}$. We compute the second derivative of the interaction potential at its minimum with respect to the respective mesons $\varphi_i$,
\begin{align}
m^{*}{_{\varphi_{ij}}^2} &=\lim_{\varphi \rightarrow \langle \varphi \rangle}\frac{\partial^2}{\partial \varphi_i\partial \varphi_j} U \,,
\label{eq:meson_mass_formula}
\end{align}
with $\varphi_{i} =\pi, \eta, \eta^{\prime}, K, {\omega}, \rho, K^*, {\phi}$, and for the vacuum expectation for the mesons we consider $\left<\varphi\right>=0$. 
Since the derivative in \Cref{eq:meson_mass_formula} is with respect to pseudoscalar and vector mesons, only the $\mathcal{L}^u_{\rm esb}$ and $\mathcal{L}_{\rm {vec }}$ will contribute to the in-medium masses. It is also noted that there are no cross terms between pseudoscalar and vector mesons. 

\subsubsection{Pseudoscalar mesons}
\label{sec:in-medium_PS_mesons}

Spontaneous chiral symmetry breaking leads to the emergence of Goldstone bosons that are massless. Since chiral symmetry is only an approximate symmetry, the pion is expected to have a finite mass, which remains relatively small compared to other hadrons~\cite{Koch:1997ei}. This is done by introducing a explicit symmetry term breaking which allows to reproduce the realistic mass terms of the pseudoscalar mesons.  The explicit chiral symmetry-breaking Lagrangian term in the CMF model is written as
\begin{align}
\mathcal{L}^u_{\rm esb}&=\frac{\chi^2}{\chi_0^2}\bigg(-\frac{1}{2}m^2_{\eta^0}\mathrm{Tr}Y^2 \nonumber \\
&-\frac{1}{2}\mathrm{Tr}\left[A_p\left(u (X+iY) u+u^\dagger (X-iY) u^\dagger\right)\right]\bigg)\,,
\label{eq:L_esb}
\end{align}
where $X$ is the scalar meson nonet (\Cref{eq:X_matrix}), $Y$ is the pseudoscalar singlet (\Cref{eq:Y_matrix}), and $\chi$ is the dilaton field with a vacuum value of $\chi_{0}$. Moreover, $u$ is the chiral transformation operator given by
\begin{equation}
    u=e^{\frac{i}{2\sigma_{0}}\pi^{a}(x)\lambda_{a}\gamma_{5}}=e^{\frac{i}{\sqrt{2}\sigma_{0}}P\gamma_{5}}\,,
\end{equation}
where $P=\frac{\pi^{a}\lambda_{a}}{\sqrt{2}}$ is the pseudoscalar octet matrix (\Cref{eq:PS_matrix}), $\lambda_{a}$ are the Gell-Mann matrices, and $\gamma_{5}$ is the fifth Dirac gamma matrix. 
The matrix of explicit symmetry-breaking parameters is given by  $A_p = \frac{1}{\sqrt{2}} \mathrm{diag}(m_\pi^2 f_\pi, \, m_\pi^2 f_\pi, \, 2m_K^2 f_K - m_\pi^2 f_\pi)$ ~\cite{Koch:1997ei}, where  $f_\pi$  and  $f_K$ are the decay constants of pions and kaons, respectively. This term generates a pion mass and establishes partially conserved axial current (PCAC) relations for the  $\pi$  and  $K$  mesons. The power selected for the dilaton field aligns with the dimensionality of the chiral condensate fields~\cite{Papazoglou:1997uw}. We assume the frozen glueball limit, $\chi$=$\chi_0$~\cite{Cruz-Camacho:2024odu}. 

The pseudoscalar meson fields in $u$ serve as the angular parameters of the chiral transformation in the non-linear realization of the SU(3) sigma model. In \Cref{eq:L_esb}, the first term, which breaks the $U(1)_A$ symmetry, gives the mass to the pseudoscalar singlet. The second term is motivated by the explicit symmetry-breaking term of the linear sigma model~\cite{Papazoglou:1998vr}. Now using \Cref{eq:meson_mass_formula} and expanding $u$ up to second order in the pseudoscalar meson fields (or up to second order in $P$) in $\mathcal{L}^u_{\rm esb}$, we can derive the in-medium masses of pseudoscalar mesons:
\begin{itemize}
    \item Pions:
    
    \begin{align}  
{m^{*^2}_{\pi^0/\pi^+/\pi^-}}=m^2_{\pi}\frac{\sigma}{\sigma_0}\,,
\label{eq:non_deg_pi_mass}
\end{align}
\item Kaons:
\begin{align} 
{m^{*^2}_{K^+/K^-}}= \frac{0.5 m^2_{K}  \left(2 \zeta + \sqrt{2} \left(\delta + \sigma\right)\right) \left(\sqrt{2} \sigma_{0} + 2 \zeta_{0}\right)}{\left(\sigma_{0} + \sqrt{2} \zeta_{0}\right)^{2}}\,,
\label{eq:non_deg_Kpm_mass}
\end{align}
\begin{align} 
{m^{*^2}_{K^0/\bar K^0}}=\frac{0.5  m^2_{K}  \left(2 \zeta + \sqrt{2} \left(-\delta + \sigma\right)\right) \left(\sqrt{2} \sigma_{0} + 2 \zeta_{0}\right)}{\left(\sigma_{0} + \sqrt{2} \zeta_{0}\right)^{2}}\,,
\label{eq:non_deg_K0b_mass}
\end{align}
\item $\eta^8$:
\begin{align}  
{m^{*^2}_{\eta^8}}=\frac{m^2_{\pi} \sigma \sigma_{0} +  \sqrt{2} \zeta \left(\sqrt{2} m^2_{K} \left(\sqrt{2} \sigma_{0} + 2 \zeta_{0}\right) - 2 m^2_{\pi} \sigma_{0}\right)}{\sigma_{0}^{2} + 4 \zeta_{0}^{2}}\,,
\label{eq:no_mix_eta8_mass}
\end{align}
\item $\eta^0$:
\begin{align}
{m^{*^2}_{\eta^0}}=m^2_{\eta^0}\,.
\label{eq:eta0_mass}
\end{align}
\end{itemize}

We note that the masses in  Eq.~\Cref{eq:non_deg_pi_mass} for the different pions are the same, and they do not depend on the $\delta$ mean-field. On the other hand, the masses of the charged kaon doublet (\Cref{eq:non_deg_Kpm_mass}) are the same but depend on the $\delta$ field, and the masses of the uncharged kaon doublet (\Cref{eq:non_deg_K0b_mass}) are also the same but with opposite sign of the $\delta$ field. However, in the isospin-symmetric matter ($\mu_Q=0$) case that we study in this work,  the $\delta$ mean-field remains zero, therefore, the masses of all kaons become equal.  In \Cref{eq:eta0_mass}, we find that the mass of the pseudoscalar singlet $\eta^0$ meson is not medium dependent, as obtained from the first term in \Cref{eq:L_esb}, responsible for breaking the  $U(1)_A$  symmetry  and therefore we put it to the vacuum value, ${m_{\eta^0}}$=958 MeV}.

Using the above expressions for meson masses in the vacuum, i.e., $\sigma\rightarrow\sigma_0,$  and $\zeta\rightarrow\zeta_0$, we reproduce the vacuum masses of respective mesons, where  $m_\pi$=139 MeV and  $m_K$=498 MeV,  and using them further in \Cref{eq:no_mix_eta8_mass}, we get $m_{\eta^8}$=574.74 MeV. Moreover, using \Cref{eq:non_deg_pi_mass,eq:non_deg_Kpm_mass,eq:no_mix_eta8_mass}, we also obtain the following relation
\begin{align}  
3{m^2_{\eta^8}} +1.08 m^2_{\pi}-4.08 m^2_{K}=0\,,
\label{eq:eta_mass}
\end{align}
which approximately gives the Gell-Mann--Okubo mass formula \cite{Papazoglou:1997uw}
\begin{align}  
3{m^2_{\eta}} + m^2_{\pi}- 4m^2_{K}=0\,.
\end{align}

 This small discrepancy is due to the fact that  in the CMF model with the non-linear realization of chiral symmetry, there is no mixing between $\eta^8$ and $\eta^0$ mesons, therefore the model does not accurately reproduce the vacuum mass of the physical $\eta$ meson i.e.  $m_\eta=547.86$ MeV. For simplification, we use from now on $\eta$ for $\eta^8$ and $\eta^\prime$ for $\eta^0$ throughout the paper.

We also checked that in the chirally restored phase ($\sigma\rightarrow0$ and $\zeta \rightarrow0$) the mass of the pseudoscalar mesons goes to zero
\begin{align*}  
{m^*_{\pi}}= {m^*_{K}}={m^*_{\eta}}=0\,.
\end{align*}
The exception is ${m^*_{\eta^\prime}}$ (\Cref{eq:eta0_mass}),  as it does not depend upon scalar fields and is, therefore, constant. This is a consequence of $\eta^\prime$ being a pseudoscalar singlet, which is not one of the pseudoscalar mesons that act as the parameters of the symmetry transformation $u$.

\subsubsection{Vector mesons}
\label{sec:in-medium_V_mesons}

Recently, in Ref.~\cite{Kumar:2024owe}, we have broken the vector meson mass degeneracy in the CMF model by adding a cross term between scalar and vector meson mean fields to the Lagrangian, resulting in a redefinition of the vector fields. The total field-redefined vector interaction Lagrangian term (without the kinetic term) for the C4 coupling scheme\footnote{There are three main different chiral invariant possibilities for the vector
self-interaction terms C1, C3, and C4, plus two more that were never studied in detailed because they did not seem to produce physical results. All other chiral invariants can be derived from these couplings~\cite{Dexheimer:2015qha,Malik:2024qjw}.} which results in better agreement with the measure maximum mass of neutron stars~\cite{Dexheimer:2015qha,Malik:2024qjw}, is given by
\begin{align}
\mathcal{ L}_{\rm  vec}&= \mathcal{ L}_{\rm  vec}^{\rm  m} +\mathcal{ L}_{\rm  vec}^{\rm SI}\,,
\label{eq:L_vec_net}
\end{align}
where the mass term is given by
\begin{align}
\mathcal{ L}_{\rm  vec}^{\rm  m}=\frac{1}{2} \left(m_{\rho}^2 {\rho}^2+m^2_{K^*} K{^*}^2 +m_{\omega}^2 {\omega}^2+m_{\phi}^2 {\phi}^2\right),
\label{eq:L_vec_mass}
\end{align}
and the self-interaction term for the C4 vector coupling scheme is given by~\cite{Kumar:2024owe}
\begin{align}
\mathcal{L}_{\rm  vec}^{\rm SI}&=g_4\left({\omega}^{4}+2\sqrt{2}\bigg(\frac{Z_{\phi}}{Z_{\omega}} \bigg)^{1/2} {\omega}^{3} {\phi}+3 \bigg(\frac{Z_{\phi} }{Z_{\omega}}\bigg)  {\omega}^{2} {\phi}^{2}\right. \nonumber \\
&\left.+\sqrt{2} \bigg(\frac{Z_{\phi}}{Z_{\omega}}\bigg)^{3/2}{\omega} {\phi}^{3}+\frac{1}{4} \bigg(\frac{Z_{\phi} }{Z_{\omega}}\bigg)^2{\phi}^{4}\right)\,,
\end{align}
where $Z_{\phi}=\frac{1}{\left(1-\upmu \zeta_0^2\right)}$ and $Z_{\omega}=\frac{1}{\left(1-\upmu \frac{\sigma_0^2}{2}\right)}$ are field-redefined constants with $\upmu=$ $0.41/\sigma_0^2$ being a parameter fitted to obtain  the correct vector meson masses~\cite{Kumar:2024owe}. Now, using \Cref{eq:meson_mass_formula},  the in-medium masses of vector mesons are calculated as
\begin{itemize}
\item ${\omega}$:
\begin{align}  
m^*{_{{\omega}}^{2}}&=m_{{\omega}}^{2}+6 g_4 \bigg(\frac{Z_{\phi} }{Z_{\omega}}\bigg)  {\phi}^2\,,
\label{eq:omega_mass}
\end{align}
\item ${\phi}$:
\begin{align}  
m^*{_{{\phi}}^{2}}&= m_{{\phi}}^{2}+6g_4  \bigg(\frac{Z_{\phi} }{Z_{\omega}}\bigg){\omega}^2\,,
\label{eq:phi_mass}
\end{align}
\item $\rho$:
\begin{align}  
m^*{_{\rho}^{2}}=m_{\rho}^{2}\,,
\label{eq:rho_mass}
\end{align}
\item $K^*$:
\begin{align}  
m^*{_{K^*}^{2}}=m_{K^*}^{2}\,.
\label{eq:Kstar_mass}
\end{align}
\end{itemize}

In general, the mass of the $\rho$ meson should  also depend on the medium, but in the C4 coupling scheme the $\rho$ meson does not appear in the self-interaction term, which results in the
$\rho$ meson mass remaining constant~\cite{Kumar:2024owe}. On the other hand, since the $K^*$ meson is off-diagonal in the vector meson matrix (\Cref{eq:V_matrix}), it does not contribute to the trace when forming chiral invariants~\cite{Kumar:2024owe} under chiral transformations. As a result, its mass also remains constant.

Finally, in the equations of motion (\Cref{eq:eom_sig,eq:eom_zet,eq:eom_del,eq:eom_ome,eq:eom_rho,eq:eom_phi}), we see that the derivatives of meson fields with respect to scalar and vector fields are needed. We have calculated these derivatives for the first time in the CMF model. They are shown in the \Cref{sec:derivative_mass_fields} of this paper.

\section{Results}
\label{sec:results}
In this section, we present our numerical results on the impact of interacting thermal mesons with mean-field dependent masses on the thermodynamics of matter within a large range of $T$ and $\mu_B$.  We consider the baryon octet and decuplet, pseudoscalar, and vector mesons (with field-redefined C4 vector coupling) as degrees of freedom in what we call the new hadronic CMF model with improved meson description (mCMF model) across various combinations of $\mu_B$ and $T$. To assess the significance of incorporating interacting thermal mesons, we compare the results of the mCMF model with those from the standard CMF framework, where thermal mesons are non-interacting and their in-medium masses remain unchanged with varying density and temperature. Finally, we contrast these findings with those from a CMF framework that excludes thermal mesons altogether (CMF noTM), providing a comprehensive analysis of how different mesonic treatments can influence the thermodynamics of strongly interacting matter. Note that because of isospin symmetry, the masses of isospin partners are degenerate, e.g., pions, and kaons have the same mass, i.e. $m_{\pi^{0}}=m_{\pi^{+}}=m_{\pi^{-}}$, $m_{K^{0}}=m_{\bar{K}^{0}}$, and $m_{K^{+}}=m_{K^{-}}$.

In the following subsections, we begin with displaying the results for meson mean fields (from hereon simply mean fields) and in-medium masses of interacting thermal mesons. Next, we present population distributions for individual baryon and meson species concerning $\mu_B$ and $T$. We then explore selected thermodynamic variables calculated within the mCMF model and compare these with the equation of state from lattice QCD extrapolated to finite $\mu_B$ and both ideal HRG and van der Waals HRG models. Additionally, we examine the partial pressures of various particles from the mCMF model categorized by their baryon and strangeness numbers at $\mu_B/T=0$, comparing these with lattice QCD results. Finally, we present the mCMF-derived second-order baryon susceptibility and compare it with both the  van der Waals HRG and lattice QCD results at $\mu_B=0$.

\begin{figure*}
    \centering
    \includegraphics[scale=0.55]{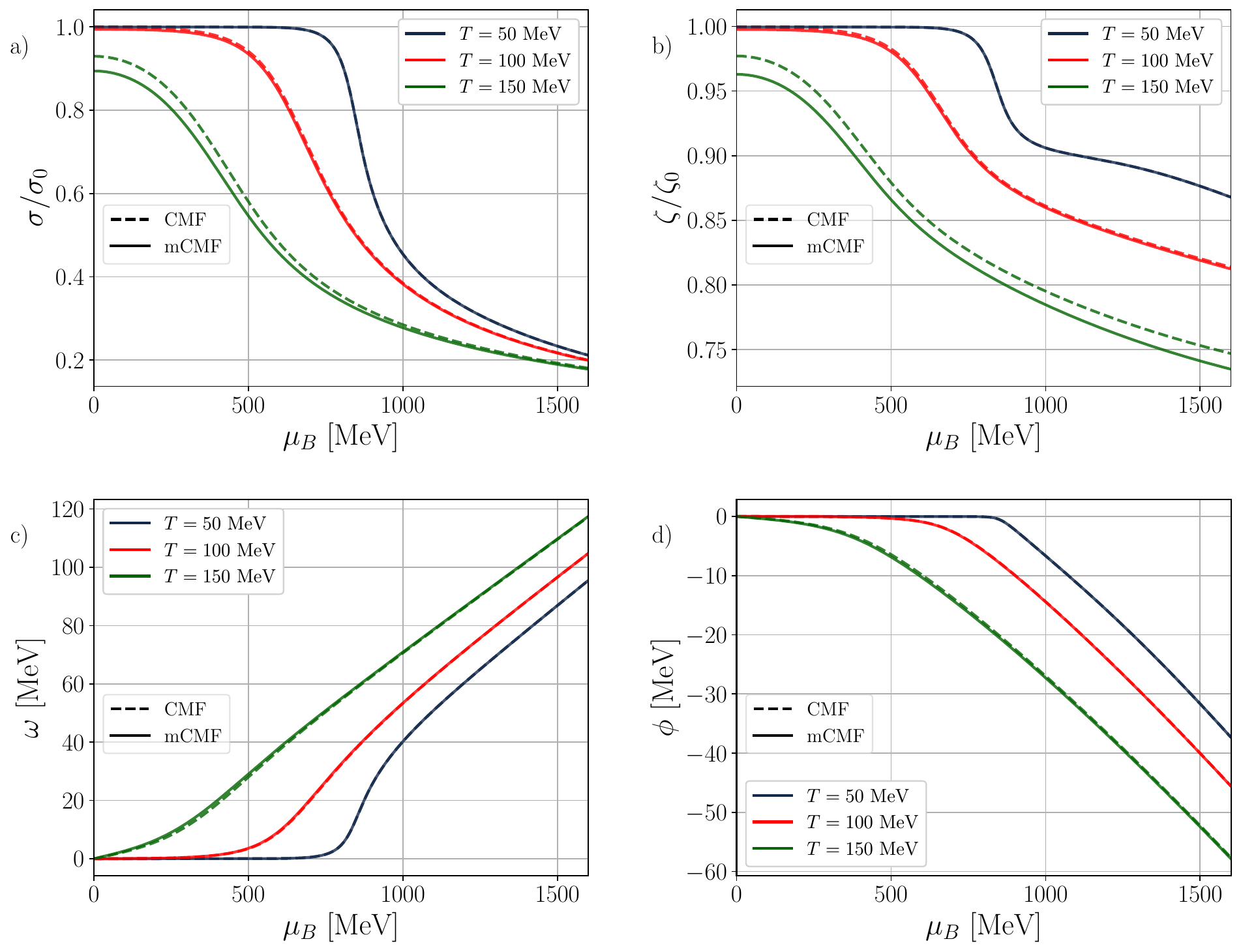}
    \caption{Meson mean-fields vs. baryon chemical potential. The panels are a)  scalar $\sigma$ field normalized by vacuum value, b) scalar $\zeta$ field normalized by vacuum value c) vector ${\omega}$ field, and d) vector ${\phi}$ field. A comparison of results for different values of temperature, each for non-interacting thermal mesons (CMF) and interacting thermal mesons (mCMF), is shown.}
\label{fig:mean_fields_C4_octet_and_decuplet_hadrons_cpb}
\end{figure*}

\subsection{Interacting thermal meson contribution to the mean-fields}
\label{sec:mean-fields}

\begin{figure*}
    \centering
    \includegraphics[scale=0.52]{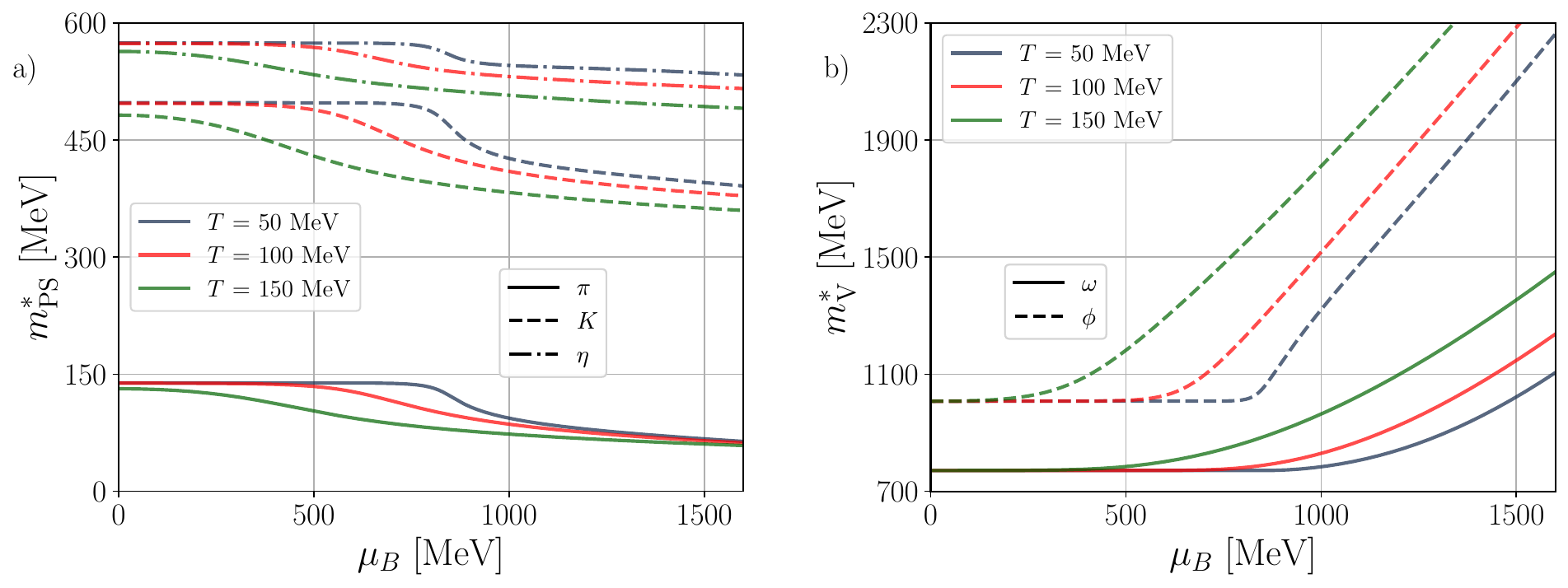}
    \caption{In-medium masses of mesons vs. baryon chemical potential for the interacting meson (mCMF) case. Panel a) shows pseudoscalar mesons, whereas panel b) shows vector mesons. A comparison of results for different values of temperature is shown.}
\label{fig:effective_meson_mass_C4_octet_and_decuplet_hadrons_cpb}
\end{figure*}

In~\Cref{fig:mean_fields_C4_octet_and_decuplet_hadrons_cpb}, we examine the dependence of the mean fields on $\mu_B$ for several temperatures. The fields analyzed include the scalar $\sigma$ and $\zeta$ (with hidden strangeness) fields normalized by their respective vacuum values and the vector $\omega$ and $\phi$ (with hidden strangeness) fields. Results are shown for two cases: non-interacting mesons, where thermal meson contributions are not medium-dependent; and interacting mesons, where thermal meson interactions are included, modifying their mass (\Cref{eq:non_deg_pi_mass,eq:non_deg_Kpm_mass,eq:non_deg_K0b_mass,eq:no_mix_eta8_mass,eq:omega_mass,eq:phi_mass})  via the back-reaction terms in the equations of motion (\Cref{eq:eom_sig,eq:eom_zet,eq:eom_del,eq:eom_ome,eq:eom_rho,eq:eom_phi}). The isovector meson fields $\delta$ and $\rho$ remain zero in the isospin-symmetric case ($\mu_Q=0$), which is what we study in this work.

In panel (a) of \Cref{fig:mean_fields_C4_octet_and_decuplet_hadrons_cpb}, we show the $\sigma$ mean-field which decreases smoothly with increasing $\mu_B$. This decrease is influenced by both $T$ and the inclusion of in-medium effects. The interacting meson (mCMF) cases display an earlier and slightly less steep decline compared to non-interacting meson (CMF) cases. Higher $T$ further accentuates these effects, suggesting stronger medium modifications to the $\sigma$ field. 

Panel (b) of \Cref{fig:mean_fields_C4_octet_and_decuplet_hadrons_cpb} presents the $\zeta$ mean-field, which shows a subtler dependence on $\mu_B$ compared to the $\sigma$ field (notice see different y-axis scale). The mCMF and CMF cases for $\zeta$ reveal small deviations, with the mCMF configurations experiencing a marginally more pronounced reduction at higher $T$ (especially at large $\mu_B$), reflecting the influence of thermal meson feedback. Since chiral symmetry breaking is associated with the emergence of scalar condensates, it can be employed as order parameters in chiral models. The decreasing trend in both scalar condensates as a function of $\mu_B$ signals the restoration of chiral symmetry. In particular, the $\sigma$ field (a proxy for the chiral condensate associated with light quarks) shows a smooth crossover behavior with respect to $\mu_{B}$ in the hadronic CMF and mCMF models. Its inflection point (or where the $\mu_{B}$ derivative peaks) provides the pseudocritical $\mu_B$, which at $T=50$ MeV is approximately $\mu_{B}\approx800$ MeV. This inflection point or pseudotransition point moves to a smaller $\mu_{B}$ value as $T$ increases. As $T$ decreases the inflection point moves in the opposite direction (to a higher $\mu_{B}$ value), but the phase transition never becomes first order.

The behavior of the $\omega$ mean-field, depicted in panel (c) of \Cref{fig:mean_fields_C4_octet_and_decuplet_hadrons_cpb}, shows a noticeable increase with increasing $\mu_B$, illustrating the anticipated strengthening of the repulsive vector potential in dense nuclear matter. Here, the difference between CMF and mCMF cases can only be identified at higher $T$, with the mCMF cases showing slightly higher $\omega$ values, indicating that thermal meson interactions contribute additional repulsion in the system. Moreover, in panel (d) of \Cref{fig:mean_fields_C4_octet_and_decuplet_hadrons_cpb}, the $\phi$ field, which primarily impacts strange particles, remains zero across the low $\mu_B$  until the emergence of strange baryons (also see \Cref{fig:population_C4_octet_and_decuplet_hadrons_T100_cpb,fig:population_C4_octet_and_decuplet_hadrons_T150_cpb} and the corresponding discussion), but decreases with increasing $\mu_B$, particularly for higher $T$. Notably, mCMF cases produce slightly more negative values for $\phi$ than CMF, suggesting that thermal meson interactions modify repulsion in the strange sector.

In a hadronic medium, at low temperatures the dominant contribution to the EoS comes from the pseudoscalar mesons due to their light mass than the vector mesons. The dependence of the pseudoscalar meson in-medium masses on the scalar fields (\Cref{eq:non_deg_pi_mass,eq:non_deg_Kpm_mass,eq:non_deg_K0b_mass,eq:no_mix_eta8_mass}) emphasizes the role of the scalar meson fields in determining the medium modifications observed in the results. The vector meson fields ${\omega}$ and ${\phi}$ do not feed back to the pseudoscalar in-medium mass, therefore the vector fields remain largely insensitive to the in-medium modifications, with only minor differences emerging  at high $T$. This behavior supports the understanding that the medium's scalar meson mean fields have a stronger sensitivity to $T$ effects than the vector components in this model (refer to \Cref{fig:mean_fields_C4_octet_and_decuplet_hadrons_cpb}). The interacting meson (mCMF) case, incorporating meson feedback, presents a model closer to real nuclear matter conditions, where pseudoscalar mesonic interactions influence the overall dynamics, especially at higher $T$. 

\begin{figure*}[t!]
    \centering
    \includegraphics[scale=0.55]{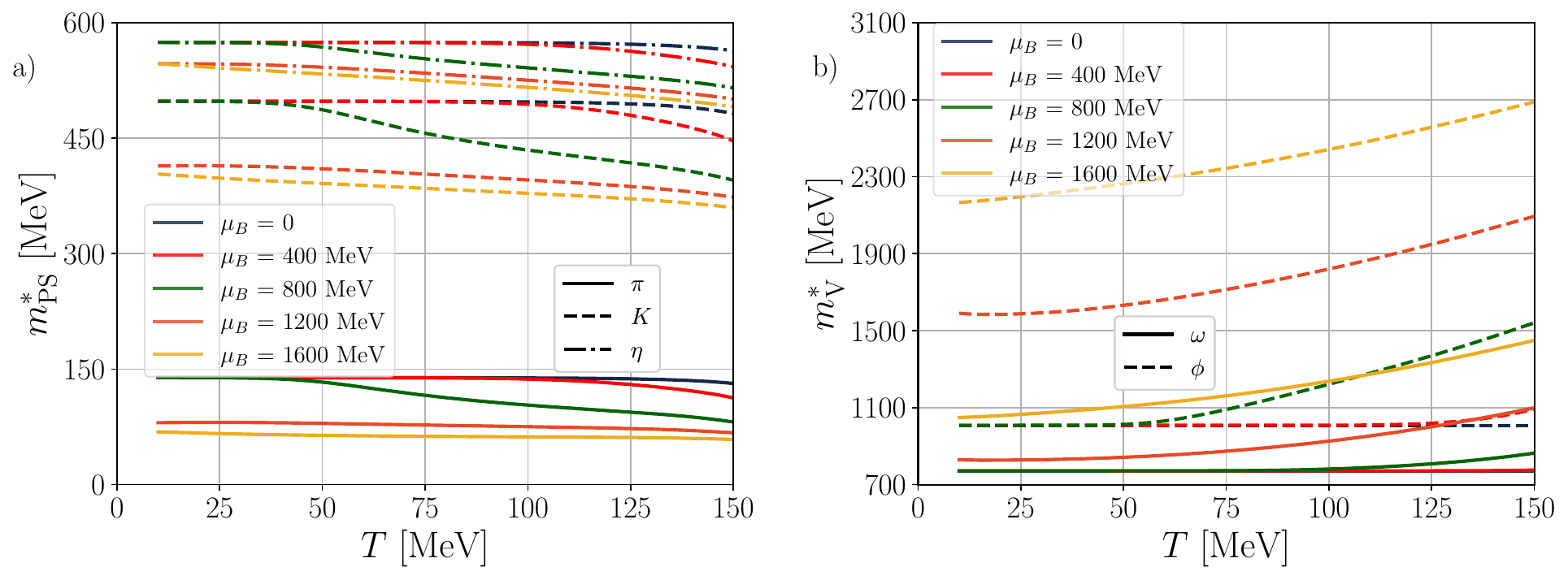}
    \caption{In-medium masses of mesons vs. temperature for the interacting meson (mCMF) case. Panel a) shows pseudoscalar mesons, whereas panel b) shows vector mesons. A comparison of results for different values of baryon chemical potential is shown.}
\label{fig:effective_meson_mass_C4_octet_and_decuplet_hadrons_temp}
\end{figure*}

\subsection{Interacting thermal meson contribution to the in-medium masses of mesons}
\label{sec:in-medium_mass_mesons}

In ~\Cref{fig:effective_meson_mass_C4_octet_and_decuplet_hadrons_cpb}, we analyze the in-medium masses of pseudoscalar (panel (a)) and vector (panel (b)) mesons as functions of $\mu_B$ for the interacting meson (mCMF) case at different temperatures.
In panel (a), the in-medium masses of the pions shows a decrease with increasing $\mu_B$, indicating the sensitivity of pion masses to the dense nuclear environment. The effect of $T$ on pion masses is to smooth out the mass decrease, making it more gradual. In the same panel, we show the in-medium masses of the kaons, which also decrease similarly with $\mu_B$. However, the rate of decrease is more pronounced than that observed for pions, and kaon masses demonstrate a higher $T$ sensitivity even at moderate values of $\mu_B$. This $T$ dependence implies that the in-medium mass of kaons, due to its dependence on the strange $\zeta$ field along with $\sigma$ field, experience additional attraction in dense, hot nuclear matter. Furthermore, the $\eta$ meson mass exhibits again a gradual decrease with $\mu_B$, but with a somewhat higher $T$ sensitivity than pions. 
The decrease of the pion mass with temperature we see here contrasts with the results found, for example in Ref.~\cite{Schaefer:2008hk}, where, within the linear sigma model, the pion and its chiral partner (the sigma meson) masses become degenerate to a non-zero value. However, this linear sigma model contains the quark loop corrections as well as the model has different degrees of freedom, with a phase transition from mesons to quarks at increasing $T$. The inclusion of fermion loop corrections and quark degrees of freedom will be the topic of a future report in the CMF model.

Furthermore, panel (b) of \Cref{fig:effective_meson_mass_C4_octet_and_decuplet_hadrons_cpb} illustrates the in-medium masses of the $\omega$ meson, which show a steep increase with increasing $\mu_B$, indicative of the significant modification in dense matter. Notably, at $T = 150$ MeV, the mass of the $\omega$ meson increases more smoothly compared to lower $T$. Finally, the ${\phi}$ meson mass exhibits the most substantial increase with $\mu_B$ among the mesons studied. This behavior can be understood from the expressions for the in-medium $\omega$ and $\phi$ masses (\cref{eq:omega_mass,eq:phi_mass}). The in-medium mass of $\omega$ is influenced by the $\phi$ mean-field, while the $\phi$ in-medium mass depends on the $\omega$ mean-field. Since the $\omega$ mean-field exhibits a stronger growth with increasing $\mu_B$ (\Cref{fig:mean_fields_C4_octet_and_decuplet_hadrons_cpb}), this leads to a significant increase in the $\phi$ mass.

In~\Cref{fig:effective_meson_mass_C4_octet_and_decuplet_hadrons_temp}, we present the in-medium masses of pseudoscalar and vector mesons as functions of $T$ in the interacting meson (mCMF) case for different $\mu_B$s.
In panel (a), the mass of the pions shows a decrease with $T$ across all values of $\mu_B$. The effect of $\mu_B$ becomes less pronounced at higher $\mu_B$ values, where the pion mass barely changes. The in-medium masses of kaons also shows a decrease with $T$, but with a more significant dependence on $\mu_B$ than pions. This dependence suggests again that kaons experience strong medium modifications in dense baryonic environments, which are further enhanced by thermal effects. Furthermore, the $\eta$ meson mass exhibits again a gradual decrease with $\mu_B$, but with a higher $T$ sensitivity than pions for higher $\mu_B$.

The $\omega$ meson mass, shown in panel (b) of \Cref{fig:effective_meson_mass_C4_octet_and_decuplet_hadrons_temp}, exhibits a substantial increase with increasing $T$, particularly at higher $\mu_B$ values. This $T$ sensitivity reflects the strong medium effects that intensify with both $\mu_B$ and $T$. Finally, the ${\phi}$ meson mass, which is significantly influenced by $\mu_B$ and $T$, has a higher sensitivity to $\mu_B$ and $T$ in comparison with $\omega$. This behavior aligns with the fact that the in-medium mass of $\phi$ meson (\cref{eq:phi_mass}) depends on the $\omega$ mean field which grows substantially in the medium (\Cref{fig:mean_fields_C4_octet_and_decuplet_hadrons_cpb}).

\subsection{Interacting thermal meson contribution to the populations of the particles}
\label{sec:population_plot}

\begin{figure*}
    \centering
    \includegraphics[scale=0.55]{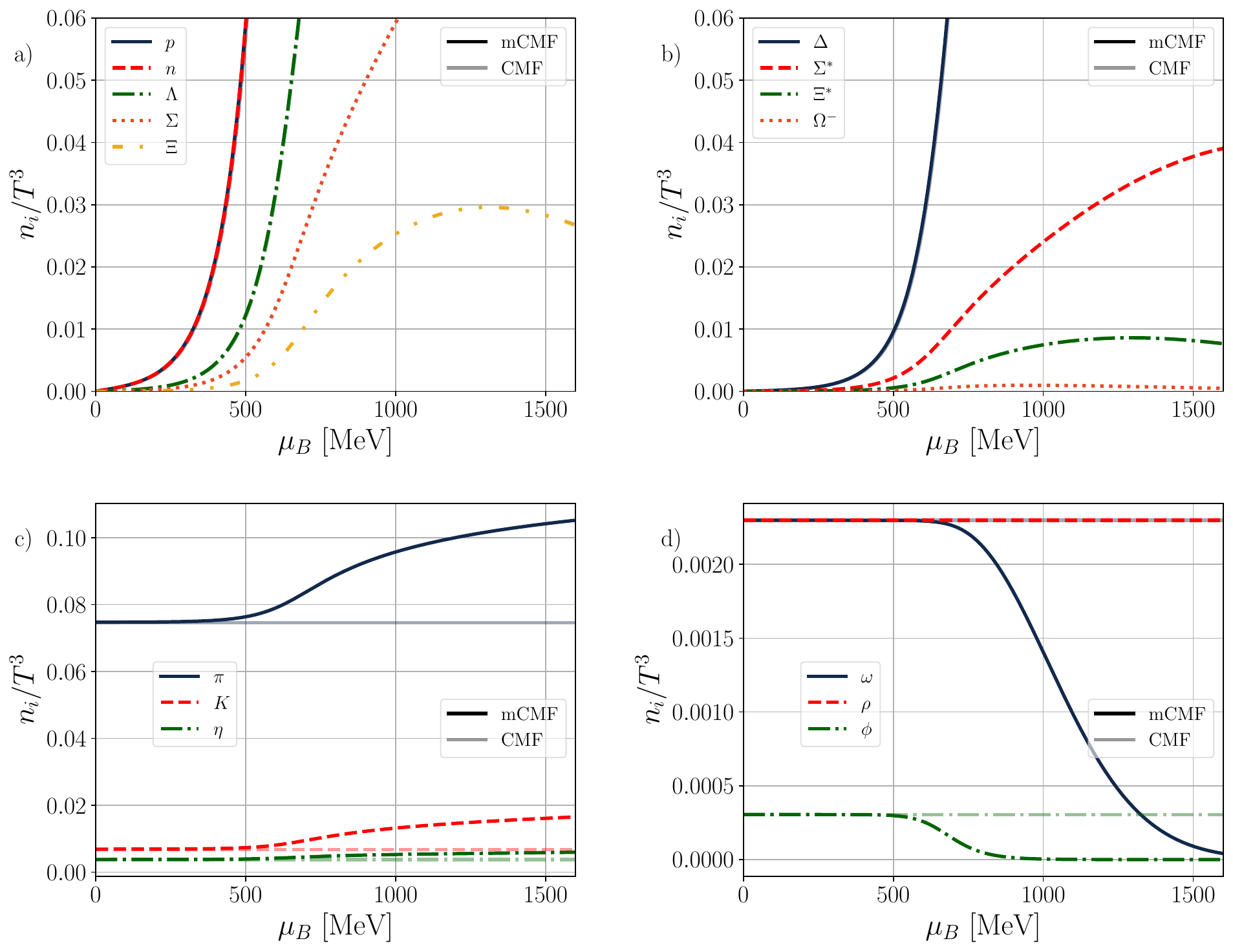}
    \caption{Scaled particle population vs. baryon chemical potential at $T=100$ MeV for the non-interacting mesons (CMF) as well as the interacting mesons (mCMF) case. The panels show number densities for the a)  baryon octet, b) baryon decuplet c) pseudoscalar mesons, and d) vector mesons. On the top panels, the curves from CMF and mCMF overlap. Note that the population of the isospin multiplets of $\Sigma$, $\Xi$, $\Delta$, $\Sigma^*$, $\Xi^*$, $\pi$, $K$ and $\rho$ overlap.}
\label{fig:population_C4_octet_and_decuplet_hadrons_T100_cpb}
\end{figure*}

In~\Cref{fig:population_C4_octet_and_decuplet_hadrons_T100_cpb}, we present the particle populations (scaled by $T^3$) as a function of $\mu_B$ at $T = 100$ MeV in the interacting meson (mCMF) and non-interacting meson (CMF) cases. 
In panel (a), for the mCMF case, we display the population of the baryon octet ($p$, $n$, $\Lambda$, $\Sigma$s, $\Xi$s) as $\mu_B$ increases. The number densities of protons and neutrons are equal due to isospin-symmetric matter and they increase monotonously with $\mu_B$. The population of hyperons, particularly $\Lambda$ and $\Sigma$ baryons, also increases, indicating that these strange baryons become more energetically favorable in the hot and dense hadronic matter. Among these two, the $\Lambda$ population increases at a larger rate because of the smaller $\Lambda$ mass than the $\Sigma$ and, most importantly, because of the more attractive $\Lambda$ interactions in the medium.  The  $\Xi$ baryons, being doubly strange and heavier, show a smaller population but still increase slightly with $\mu_B$, suggesting that they can appear at high densities. In the panel (b) of \Cref{fig:population_C4_octet_and_decuplet_hadrons_T100_cpb}, we show the densities of the baryon decuplet ($\Delta$s, $\Sigma^*$s, $\Xi^*$s, $\Omega^-$), with a relatively lower population compared to the baryon octet. Among the particles from the baryon decuplet, $\Delta$ baryons have the largest population, consistent with their non-strange content and attractive interactions in the medium. However, as $\mu_B$ increases, hyperons from the decuplet ($\Sigma^*$s, $\Xi^*$s, and $\Omega^-$) also start appearing in small fractions. Additionally, a comparison between the interacting (mCMF) and non-interacting (CMF) cases at $T$ = 100 MeV reveals no significant difference in the baryonic populations, suggesting that interactions among thermal mesons do not strongly affect baryon abundances at this temperature.
\begin{figure*}
    \centering
    \includegraphics[scale=0.55]{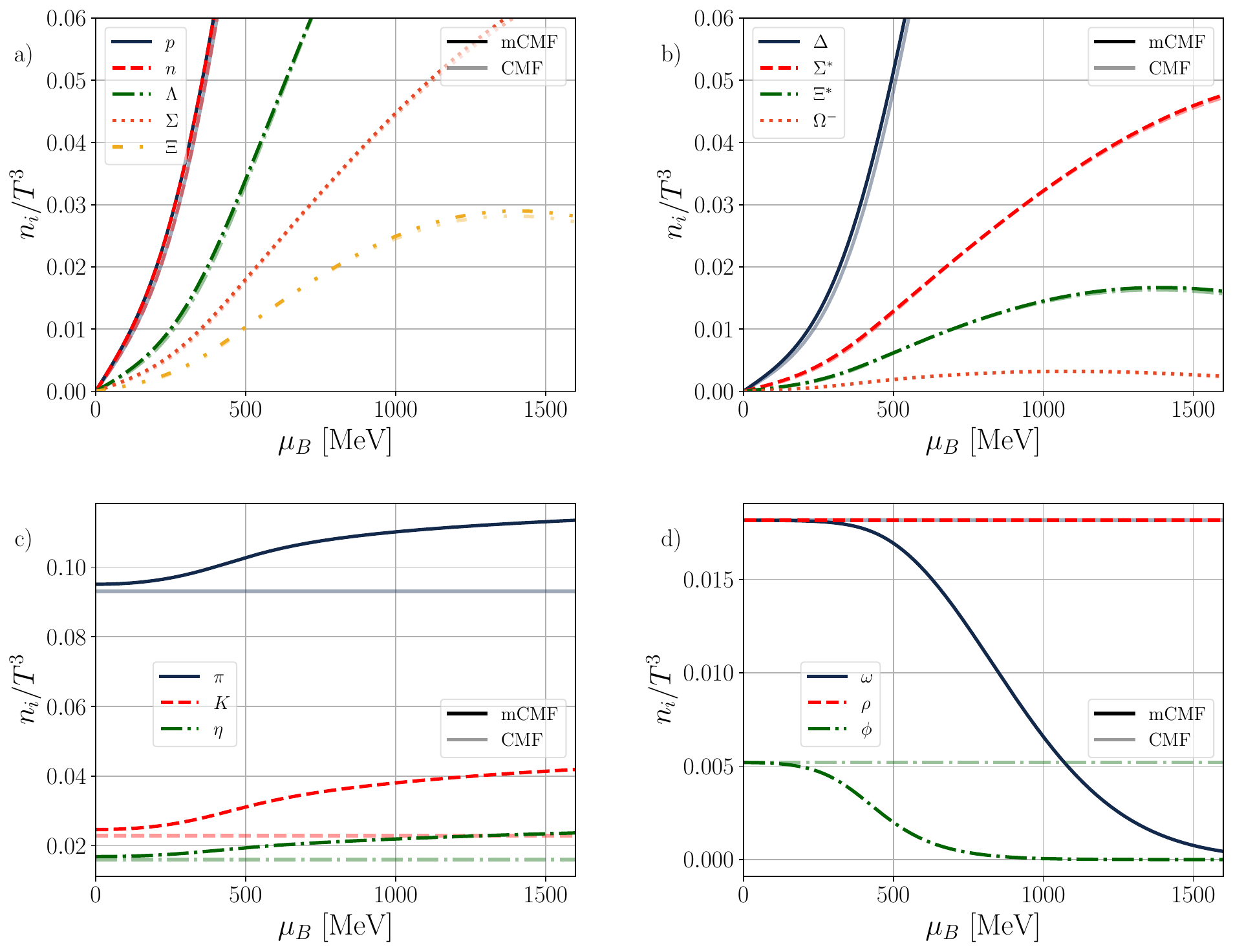}
    \caption{Scaled particle population vs. baryon chemical potential at $T=150$ MeV for the non-interacting mesons (CMF) as well as the interacting meson (mCMF) cases. The panels show number densities for the a)  baryon octet, b) baryon decuplet c) pseudoscalar mesons, and d) vector mesons.}
\label{fig:population_C4_octet_and_decuplet_hadrons_T150_cpb}
\end{figure*}

Furthermore, panel (c) of \Cref{fig:population_C4_octet_and_decuplet_hadrons_T100_cpb} illustrates the population of the pseudoscalar mesons ($\pi$, $K$, $\eta$) at $T$ = 100 MeV. Here, the pion population has a qualitatively different $\mu_B$ dependence in comparison with baryons. More kaons appear at moderate $\mu_B$ values, indicating that strange mesonic states become energetically favorable under these conditions. The $\eta$ meson population is the lowest among pseudoscalars, showing minimal change with $\mu_B$ due to its larger mass. In panel (d), the vector mesons ($\omega$, $\rho$, $\phi$) exhibit minimal populations across the $\mu_B$ range (note the different scales). The $\omega$ appears in small quantities ($n/T^3 \sim 10^{-4}$) and for mCMF it decreases with $\mu_B$ because the in-medium masses of $\omega$ increase with $\mu_B$ (\Cref{fig:effective_meson_mass_C4_octet_and_decuplet_hadrons_cpb}). The $\rho$ meson also shows a small number density ($n/T^3 \sim 10^{-4}$) and it remains constant throughout $\mu_B$ (due to its constant mass), while the $\phi$ meson has an even lower population due to its strange quark content and increased mass. The $\phi$ meson density also decreases further with $\mu_B$. Altogether, the vector mesons play a limited role in the overall particle population at $T = 100$ MeV and the given $\mu_B$ values, likely contributing more through their mean-field effects rather than direct population density. For the non-interacting meson (CMF) case, the meson masses are constant, meaning they would appear as straight horizontal lines in \Cref{fig:effective_meson_mass_C4_octet_and_decuplet_hadrons_cpb,fig:effective_meson_mass_C4_octet_and_decuplet_hadrons_temp} if they were shown. As a result, the particle populations for the non-interacting meson (CMF) case also appear as straight horizontal lines in the bottom panels of \Cref{fig:population_C4_octet_and_decuplet_hadrons_T100_cpb,fig:population_C4_octet_and_decuplet_hadrons_T150_cpb}.

\begin{figure*}
    \centering
    \includegraphics[scale=0.55]{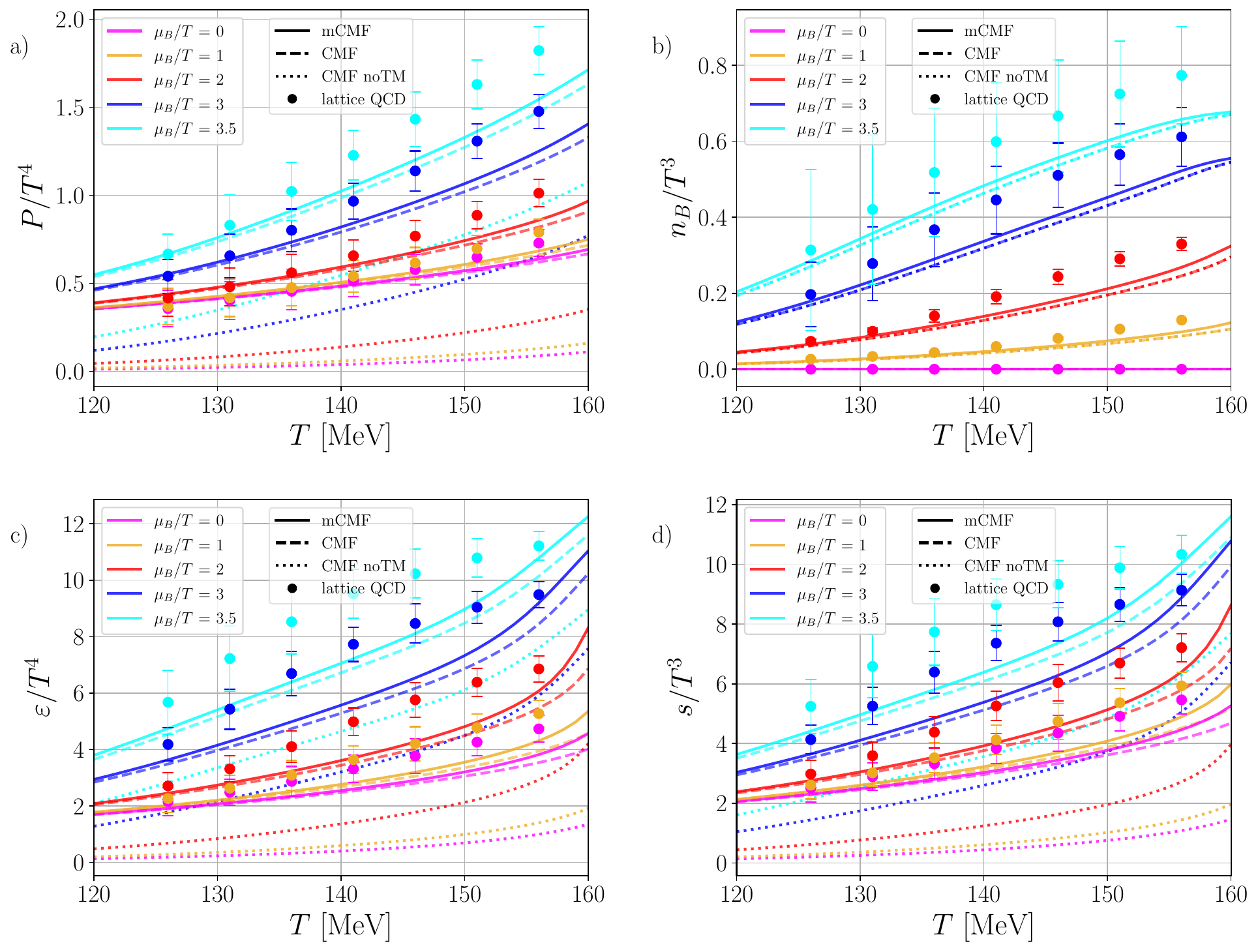}
    \caption{Thermodynamic variables scaled by appropriate powers of temperature vs. temperature for different $\mu_B/T$ ratios for the no meson (CMF noTM), non-interacting mesons (CMF), and interacting mesons (mCMF) cases. The panels are a) scaled pressure, b) scaled (net) baryon number density c) scaled energy density, and d) scaled entropy density. We also show state-of-the-art lattice QCD results extrapolated to finite $\mu_B$~\cite{Borsanyi:2021sxv} for comparison. In panel b) dotted and dashed lines overlap. }
\label{fig:observables_NM_NIM_IM_C4_octet_and_decuplet_hadrons}
\end{figure*}

In~\Cref{fig:population_C4_octet_and_decuplet_hadrons_T150_cpb}, we present the same results as in \Cref{fig:population_C4_octet_and_decuplet_hadrons_T100_cpb}, but at $T = 150$ MeV. We observe that the particle populations generally increase compared to the $T = 100$ MeV case, with a notable increase in strange particles like hyperons (from octet and decuplet) and kaons. Although vector meson densities remain minimal (panel (d)), they increase with an increase of $T$. Additionally, a comparison between the interacting (mCMF) and non-interacting (CMF) cases at  T = 150  MeV reveals small deviations with the mCMF case now exhibiting a slightly more pronounced increase in baryon populations. This trend reflects the influence of thermal meson feedback, which becomes more noticeable as thermal effects intensify.

\subsection{Interacting thermal meson contribution to thermodynamics}
\label{sec:thermodynamics_plot}

\begin{figure*}
    \centering
    \includegraphics[scale=0.55]{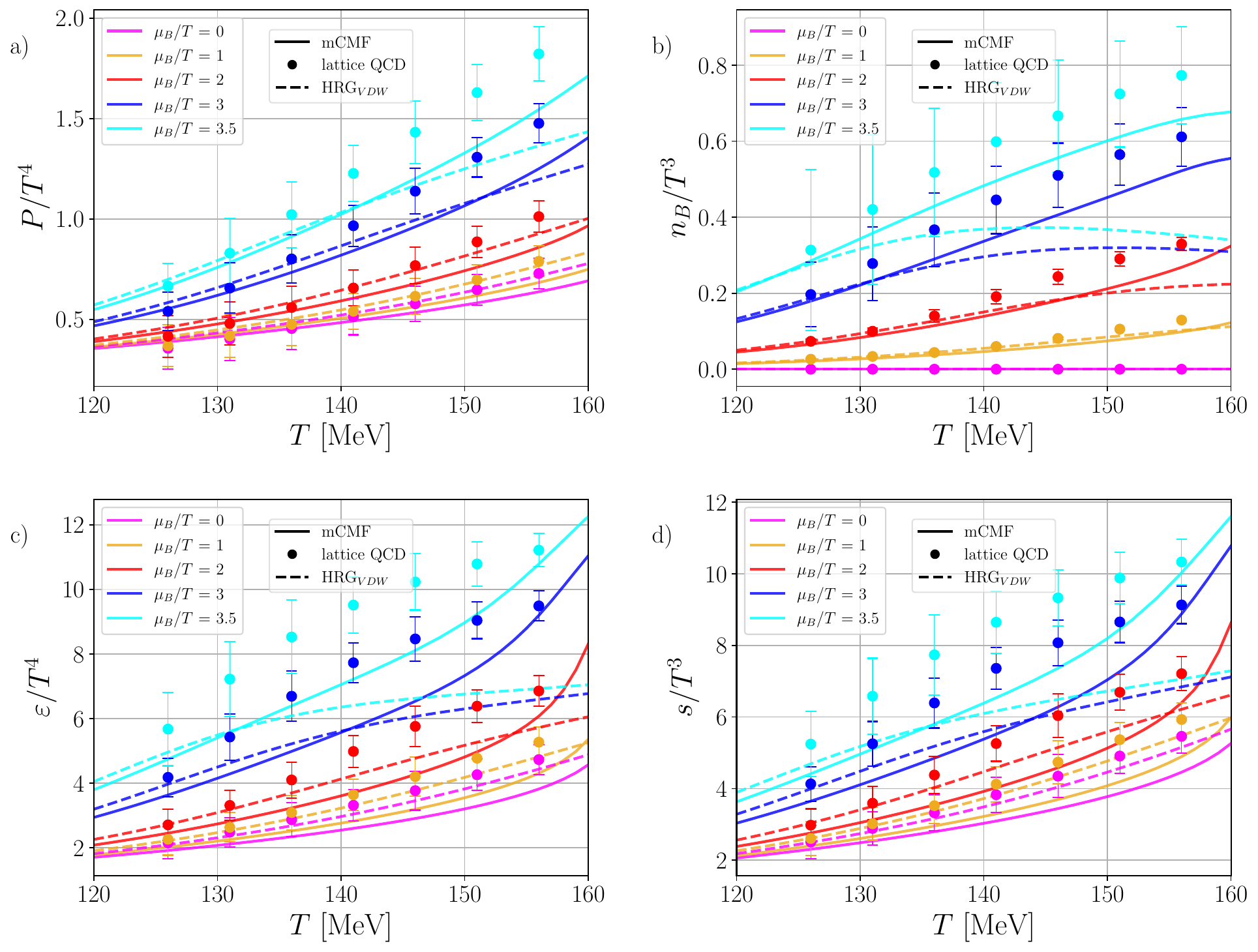}
    \caption{Thermodynamic variables scaled by appropriate powers of temperature vs. temperature for different $\mu_B/T$ ratios for only the interacting meson (mCMF) case. The panels are a) scaled pressure, b) scaled (net) baryon number density c) scaled energy density, and d) scaled entropy density. We also show state-of-the-art lattice QCD results extrapolated to finite $\mu_B$~\cite{Borsanyi:2021sxv} and the van der Waals HRG \cite{Vovchenko:2016rkn,SanMartin:2023zhv} for comparison.}
\label{fig:observables_lattice_and_HRG_C4_octet_and_decuplet_hadrons}
\end{figure*}

In~\Cref{fig:observables_NM_NIM_IM_C4_octet_and_decuplet_hadrons}, we present relevant thermodynamic variables, each scaled by an appropriate power of $T$ (to make them dimensionless), as functions of $T$ at different $\mu_B/T$ ratios. We include results from the new mCMF model, but also from the CMF model with non-interacting thermal mesons, and the CMF model without them (CMF noTM). We see in all panels that thermal mesons are essential for reproducing the thermodynamic trends observed in lattice QCD at zero $\mu_B$ and extrapolated to finite $\mu_B$. Furthermore, we see that mCMF results are an improvement over its non-interacting version.
In panel (a), the scaled pressure, $P/T^4$, increases with $T$ across all values of $\mu_B/T$, showing a stronger increase for higher $\mu_B/T$ ratios. The mCMF model results generally align well with lattice QCD results at lower $\mu_B/T$ ratios but diverge slightly at higher $T$ as $\mu_B/T$ increases. Moreover, panel (b) shows the scaled baryon number density, $n_B/T^3$,  which also increases with $T$, with a more pronounced increase at higher $\mu_B/T$ ratios. For this variable, mCMF exhibits a closer alignment with lattice QCD results at low to moderate $T$ values, while for most $\mu_B/T$, the mCMF model slightly underpredicts $n_B/T^3$ at large $T$, suggesting missing population that carries baryon number.

Furthermore, in panel (c) of \Cref{fig:observables_NM_NIM_IM_C4_octet_and_decuplet_hadrons}, the scaled energy density $\varepsilon/T^4$ shows a similar trend, increasing with $T$ and exhibiting larger values at higher $\mu_B/T$. The mCMF model follows lattice QCD predictions well at low $\mu_B/T$, especially at low $T$. However, as $\mu_B/T$ increases, mCMF underestimates the lattice results at intermediate $T$, while there is a smaller deviation at high $T$, indicating that the model has less contributions to $\varepsilon/T^4$ in the dense regime. In panel (d), the scaled entropy density $s/T^3$ demonstrates a $T$-dependent increase that aligns within uncertainties with lattice QCD data at lower $\mu_B/T$ values, while at higher $\mu_B/T$, the mCMF model underestimates the lattice predictions at intermediate $T$. 

In order to resolve the remaining discrepancy, one should also consider the presence of quark degrees of freedom, which would contribute to the thermodynamics around and above the crossover transition $T$, that at $\mu_{B}=0$ is approximately $T\approx 158$ MeV~\cite{Borsanyi:2020fev}. The inclusion of quarks and the deconfinement potential to induce a phase transition in the model could potentially improve the agreement of the mCMF EoS with lattice QCD thermodynamics. That will be the focus of future work, since we dedicate the current manuscript to describe the improvement in the hadronic description from mCMF.

\begin{figure*}
    \centering
    \includegraphics[width=0.497\textwidth]{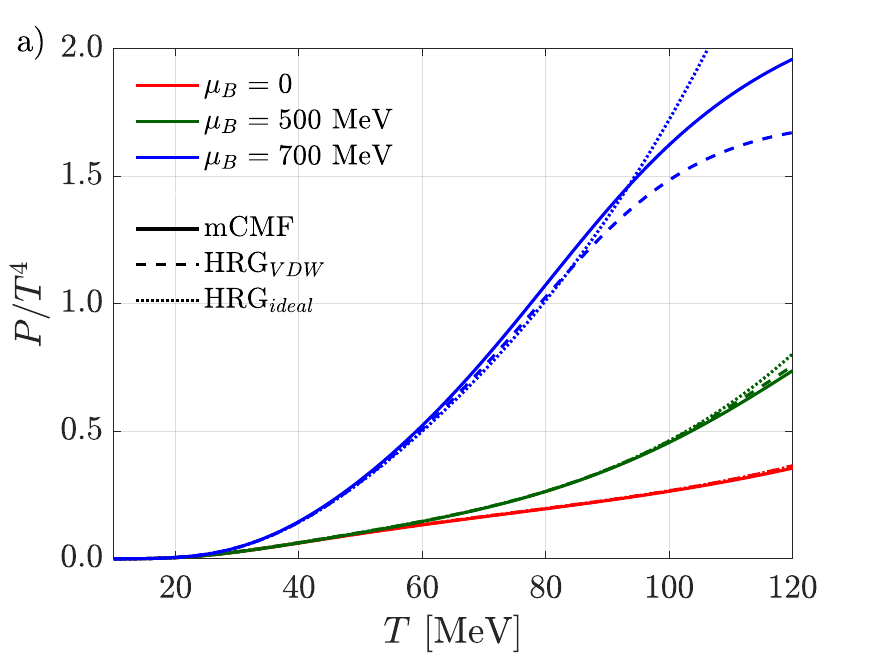}
    \includegraphics[width=0.497\textwidth]{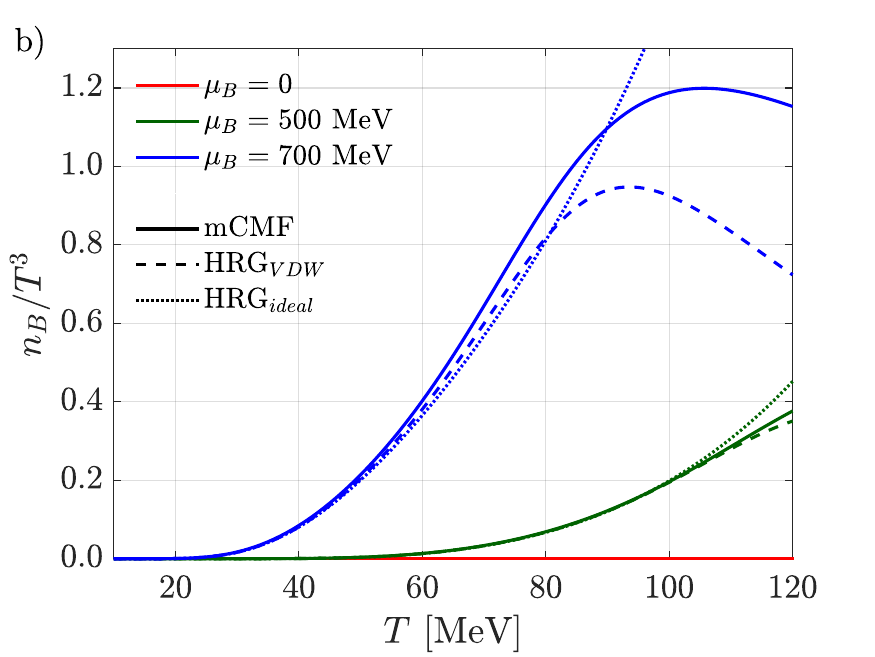}
    \includegraphics[width=0.497\textwidth]{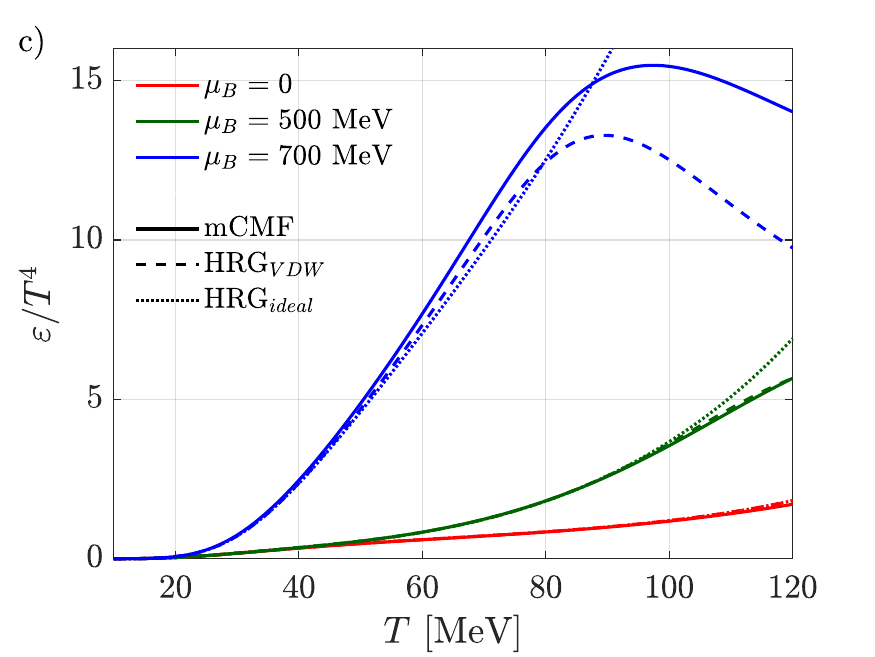}
    \includegraphics[width=0.497\textwidth]{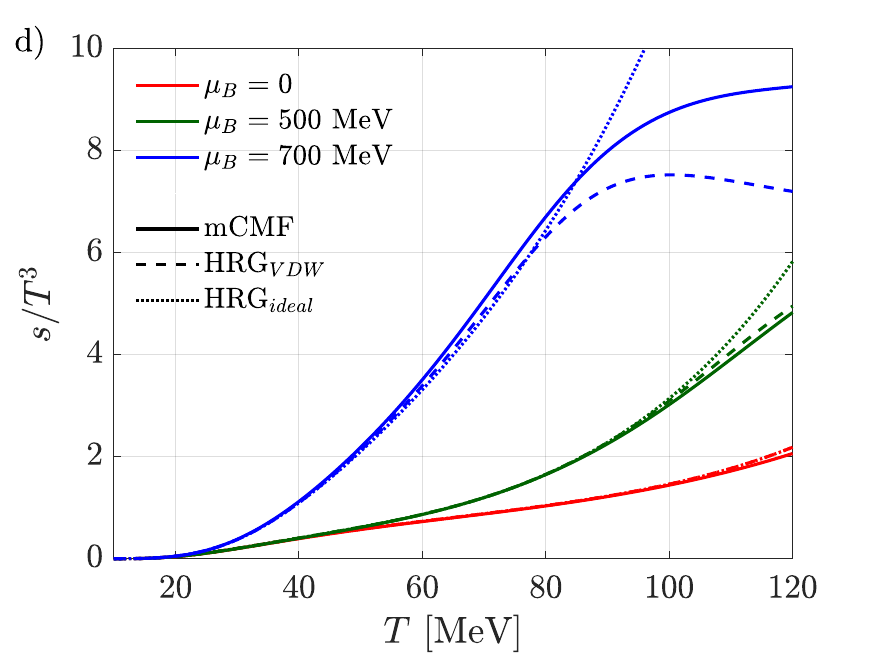}
    \caption{Thermodynamic variables vs. temperature comparing the mCMF model with the ideal Hadron Resonance Gas (HRG\textsubscript{\textit{ideal}}) and van der Waals corrected HRG (HRG\textsubscript{\textit{VDW}}) models~\cite{Vovchenko:2019pjl,SanMartin:2023zhv} for different values of $\mu_B$ focusing on the low-temperature regime. The panels display (a) scaled pressure, (b) scaled (net) baryon number density, (c) scaled energy density, and (d) scaled entropy density. }
    \label{fig:EoS_mu0_comp}
\end{figure*}

Additionally, we present our results from mCMF in \Cref{fig:observables_lattice_and_HRG_C4_octet_and_decuplet_hadrons} for the same thermodynamic quantities as the previous figure, in this case, to compare our results with the VDW HRG model contrasted with the lattice QCD results. While the scaled $P$, $n_{B}$, $\varepsilon$, and $s$ description is approximately the same at low to intermediate temperatures, the HRG$_{VDW}$ curves fall off and deviate from the lattice data at large temperatures, and even though the mCMF results underestimate lattice thermodynamics, they still follow their trend, providing a better description for hot and dense matter in this region.

\subsection{Thermodynamics comparison with HRG models}

\begin{figure*}
    \centering
    \includegraphics[scale=0.55]{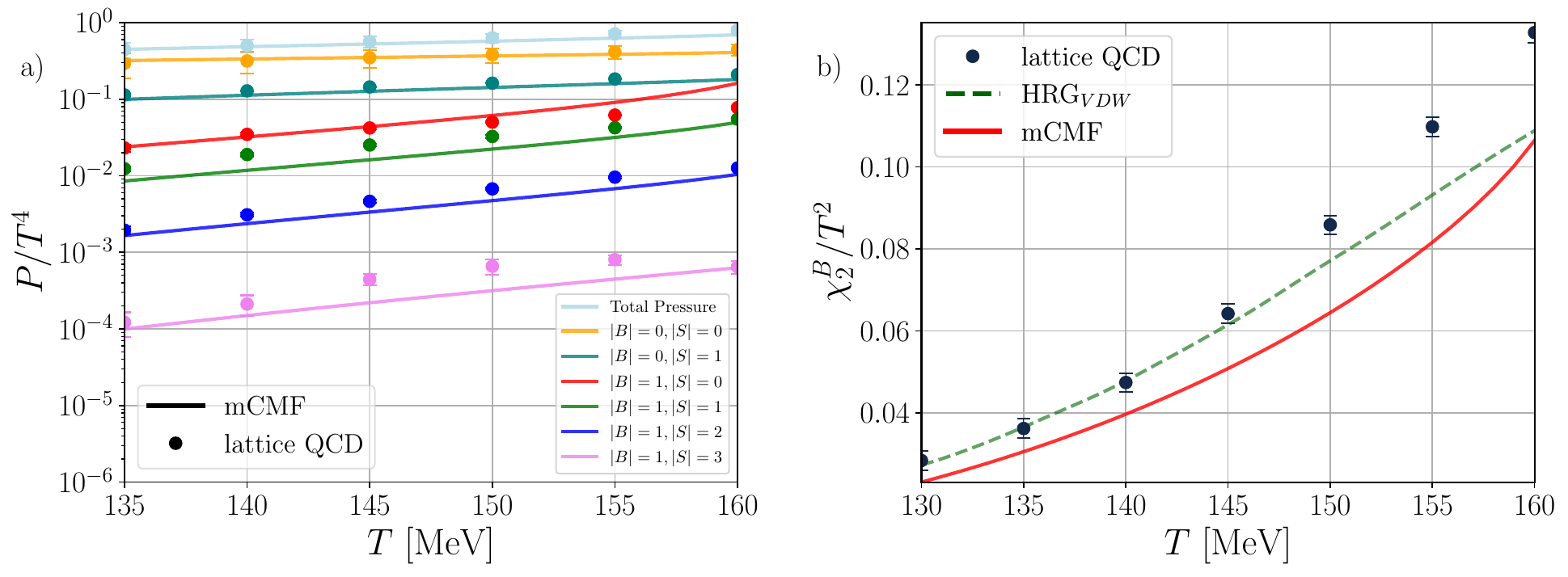}
    \caption{Thermodynamical observables vs. temperature at $\mu_B/T=0$  for the interacting mesons (mCMF) case. The panels display a) Partial pressure of contributing particles categorized in terms of baryon and strangeness number, 
    as well as total pressure. A comparison of results between mCMF and lattice QCD~\cite{Alba:2017mqu} is shown. Panel b) display second-order baryon susceptibility from mCMF, HRG$_{VDW}$~\cite{Vovchenko:2019pjl,SanMartin:2023zhv}, and lattice QCD~\cite{Borsanyi:2021sxv}.}
\label{fig:partial_pressure_chiB2_C4_octet_and_decuplet_hadrons_muBbyT0}
\end{figure*}

In~\Cref{fig:EoS_mu0_comp}, we compare the thermodynamic properties of the mCMF EoS with both the ideal HRG and the van der Waals-corrected HRG  models at lower temperatures than the ones displayed in \Cref{{fig:observables_NM_NIM_IM_C4_octet_and_decuplet_hadrons},fig:observables_lattice_and_HRG_C4_octet_and_decuplet_hadrons}. The variables are shown as functions of $T$ for different values of $\mu_B = 0$, 500, and 700 MeV. 
In panel (a) of \Cref{fig:EoS_mu0_comp}, the scaled pressure, $P/T^4$, is shown to increase with $T$ for all values of $\mu_B$. For low $T$, the mCMF model aligns closely with both HRG\textsubscript{\textit{ideal}} and HRG\textsubscript{\textit{VDW}} at all $\mu_B$'s. However, as $T$ increases, particularly for $\mu_B = 500$ MeV and $\mu_B = 700$ MeV, the mCMF model begins to deviate from the HRG models, reflecting the mCMF model's ability to be a compromise between the ideal HRG, that becomes stiffer at high $T$, and the van der Waals version that becomes softer in the same region.
Panel (b) of \Cref{fig:EoS_mu0_comp} illustrates the scaled baryon number density, $n_B/T^3$. This quantity also increases with $T$, with mCMF showing deviations from the HRG models as $T$ and $\mu_B$ increase. For the highest $\mu_B$ shown, both the mCMF model and  HRG\textsubscript{\textit{VDW}} predict a peak at $T\sim90-100$ MeV, due to the fact that their baryon density does not grow as fast as $T^{3}$. This feature is also observed for the other thermodynamic observables with their respective $T$ normalization power.

Panel (c) of \Cref{fig:EoS_mu0_comp} shows the scaled energy density, $\varepsilon/T^4$. The mCMF model once more generally agrees with the HRG models at low $T$, especially around $\mu_B = 0$. As $\mu_B$ and $T$ increase, however, there is a noticeable deviation, where the mCMF model predicts a higher energy density at intermediate $T$. This feature in mCMF helps to better describe thermal strongly interacting matter at high density, making mCMF a better suitable model to study e.g. the interior of neutron stars and their mergers. This is due to the fact that the CMF accounts for a balance of scalar-vector interactions fitted to describe saturation properties of nuclear matter. Back to panel (c) of \Cref{fig:EoS_mu0_comp}, the peak appears again for both mCMF and HRG\textsubscript{\textit{VDW}}. Finally, in panel (d) of \Cref{fig:EoS_mu0_comp}, the scaled entropy density $s/T^3$ exhibits a similar trend. Overall, our results are qualitatively closer to the HRG\textsubscript{\textit{VDW}} than HRG\textsubscript{\textit{ideal}}, indicating that introducing van der Waals corrections capture some of the interaction effects present in our approach.

\subsection{Interacting thermal meson contribution to partial pressures and susceptibility}
\label{sec:partial_pressure_chiB2}

In panel (a) of~\Cref{fig:partial_pressure_chiB2_C4_octet_and_decuplet_hadrons_muBbyT0}, we show the partial pressure contributions of different particle species within the mCMF model, scaled by $T^4$, as a function of $T$ at $\mu_B=0$. Specifically, the figure decomposes the total pressure into contributions from various particle categories, classified by their baryon number ($|B|$) and strangeness ($|S|$) content. The mCMF total pressure compares well with lattice QCD results for the temperature range under consideration. Note that the lattice QCD results are computed from combinations of susceptibilities obtained using the ideal HRG model, and assuming these relations hold for the lattice data. For the entire $T$ range analyzed, hyperons (e.g., $|B| = 1, |S| = 1$, $|B| = 1, |S| = 2$, and $|B| = 1, |S| = 3$) contribute less significantly to the total pressure, aligning with observations in lattice QCD that these multi-strange sectors span orders of magnitude smaller contributions compared to non-strange baryons~\cite{Alba:2017mqu}. The dominance of non-strange mesons  ($|B| = 0, |S| = 0$) is well-reproduced by the mCMF model, matching well with lattice data. However, as the $T$ increases, contributions from non-strange baryons ($|B| = 1, |S| = 0$) become more substantial. The mCMF model underpredicts the partial pressures in the strange meson and baryon sectors, as lattice data suggests the need for additional strange states or interactions not included in the HRG or mCMF frameworks. It is important to note that the mCMF pressure has a mean-field contribution with interacting mesons, while the HRG model pressure is based on the non-interacting ideal gas with resonances. 
The particles included in the mCMF framework, classified by their baryon and strangeness number, are the following: $\vert B \vert = 0, \vert S \vert = 0 ~ (\pi^{+}, \pi^{0}, \pi^{-}, \eta, \eta^\prime, \omega, \rho^{+}, \rho^{0}, \rho^{-}, \phi)$, $\vert B \vert = 0, \vert S \vert = 1 ~(K^{+}, K^{0}, \bar{K}^{0}, K^{-}, K^{*+}, K^{*0}, \bar{K}^{*0}, K^{*-})$, $\vert B \vert = 1, \vert S \vert = 0 ~(p, n, \Delta^{++}, \Delta^{+}, \Delta^{0}, \Delta^{-})$,  $\vert B \vert = 1, \vert S \vert = 1 ~(\Lambda, \Sigma^{+}, \Sigma^{0}, \Sigma^{-}, \Sigma^{*+}, \Sigma^{*0}, \Sigma^{*-})$, $\vert B \vert = 1, \vert S \vert = 2 ~(\Xi^{0}, \Xi^{-}, \Xi^{*0}, \Xi^{*-})$, $\vert B \vert = 1, \vert S \vert = 3 ~(\Omega^-)$.

In panel (b) of~\Cref{fig:partial_pressure_chiB2_C4_octet_and_decuplet_hadrons_muBbyT0}, we illustrate the second-order baryon susceptibility, $\chi_2^B/T^2=\partial^{2} (P/T^{4})/\partial (\mu_{B}/T)^{2}$, a key indicator of fluctuations in baryon number density and the first coefficient in the Taylor expansion of the lattice EoS at finite $\mu_{B}$. At low $T$, $\chi_2^B/T^2$ from the mCMF model aligns more closely with the HRG$_{VDW}$ and lattice QCD results, indicating a good agreement in low-$T$ baryonic response across models. However, as $T$ increases, the mCMF model predicts a slightly lower susceptibility compared to HRG$_{VDW}$ and lattice data, while still keeping the overall increasing trend of lattice QCD results.  A smaller $\chi_2^B/T^2$ at intermediate-to-high $T$ suggests again that our current hadronic mCMF is missing additional degrees of freedom and interactions, which could potentially be improved by including quark degrees of freedom and will be the topic of our subsequent work. In the case of the HRG model, these missing  interactions could be mimicked by the emergence of more massive resonances. The fact that we underestimate the lattice $\chi_{2}^{B}$ is also the reason why we underestimate the lattice EoS (4 panels in \Cref{fig:observables_lattice_and_HRG_C4_octet_and_decuplet_hadrons}) for the largest $\mu_{B}/T$ ratios at intermediate $T$.

\section{ Summary and Conclusions}
\label{sec:conclusions}
In this paper, we have introduced interactions into the thermal-meson sector of the hadronic CMF model resulting in a
a model with an improved meson description (mCMF) . 
This was achieved by evaluating in-medium modifications of the (thermal) pseudoscalar and vector mesons before applying the mean-field approximation. 
The pseudoscalar octet masses exclusively arise from the explicit chiral-symmetry  breaking term in the Lagrangian of the model, ensuring that they go to zero in the high baryon chemical potential, $\mu_B$, limit, while in-medium vector masses are affected by the self-interaction terms in the Lagrangian. We have investigated these modifications over a large range of $\mu_B$ and $T$ values but have focused on zero charge- and isospin-chemical potential, $\mu_Q=0$, relevant for relativistic heavy-ion collisions. 
For simplicity, we also kept the strangeness chemical potential zero, $\mu_S=0$.
In this regime, we have computed the equation of state and carried out detailed comparisons with two versions of the Hadron Resonance Gas model, i.e., ideal HRG and van der Waals HRG, as well as with state-of-the-art lattice QCD results extrapolated to finite $\mu_B$. We also compared mCMF to the earlier versions of the CMF model which include a non-interacting gas of thermal mesons, as well as with no thermal mesons at all. 

Our study has demonstrates that the inclusion of interacting thermal mesons impacts the behavior of scalar mean fields, with stronger medium effects observed at higher temperatures when compared to the non-interacting case. The in-medium masses of pseudoscalar mesons decrease with increasing baryon chemical potential, while vector mesons such as $\omega$ and $\phi$ show an increase, reflecting their distinct interactions with the medium, but the $\rho$ mass stays constant (as a result of our chosen parametrization more suitable to describe neutron stars).  Similarly, the pseudoscalar meson masses generally decrease with increasing temperature, whereas vector meson masses increase.

Although the baryon population is barely affected by in-medium meson effects, the pseudo-scalar and vector meson population shows substantial deviation between CMF and mCMF. This deviation increases with both $\mu_B$ and $T$. As expected, we find that pseudoscalar mesons appear in moderate quantities but are the most populated in the low baryonic chemical potential regime due to their light mass. Vector mesons populate less due to increasing mass with respect to the baryon chemical potential. At higher temperatures, strange particle populations, including hyperons and kaons, become more prominent. 

The thermodynamic variables calculated within mCMF such as pressure, number density, energy density, and entropy align reasonably well with lattice QCD results at low to moderate $\mu_B/T$ ratios. The pressure is well described within the error bars of the lattice data up to $T \simeq 150$~MeV at $\mu_B = 0$. An increase in $\mu_B/T$ results in a larger deviation from lattice results at large $T$,  possibly suggesting that excitations of additional  degrees of freedom become important at under these  conditions. 

The mCMF model description of the baryon density at given $\mu_B/T$ shows an improvement in comparison with the van der Waals-corrected HRG model. However, the energy and entropy densities are reproduced only at the lower uncertainty limit for temperatures $T \lesssim 130$~MeV under consideration. Overall, the inclusion of meson mass dependence on the in-medium mean fields together with the inclusion of the back-reaction from mesonic excitations to mean-field equations of motion brings the mCMF results closer to the lattice data in comparison with CMF (with non-interacting gas of mesons). Additionally, the mCMF model bridges the gap between ideal and van der Waals-corrected HRG models, particularly at higher temperatures and baryon chemical potentials, providing an alternative representation of thermodynamic properties, while also incorporating constraints from nuclear matter properties (including the nuclear liquid-gas phase transition critical temperature, $T^{\rm LG}_c \sim 16$ MeV) and measured maximum mass ($M_{\rm max}\sim 2 M_\odot$) of neutron stars.

In future work, the mCMF model presented here can be expanded to include quark degrees of freedom and deconfinement mechanisms, toward improving the description in the crossover regions to the deconfined matter at high temperatures and densities. Extending the formalism to include $\mu_Q\ne0$ will allow the descriptions of hot isospin-asymmetric matter, relevant for lower energy heavy-ion collisions and neutron star mergers.

\section*{Acknowledgments}
This work is partially supported by the NP3M Focused Research Hub supported by the National Science Foundation (NSF) under grant No. PHY-2116686.
We thank Hitansh Shah for providing the data for the HRG models.
VD also acknowledges support from the Department of Energy under grant DE-SC0024700.
CR also acknowledges support from the National Science Foundation under Award Number PHY-2208724, the U.S. Department of Energy, Office of Science, Office of Nuclear Physics, under Award Number DE-SC0022023, as well as by the National Aeronautics and Space Agency (NASA)  under Award Number 80NSSC24K0767. 
RR has been supported by the U.S.~NSF under grant no. PHY-2209335.

\appendix

\section{Particle multiplets}
\label{sec:appendix}

In this section, we show the particle multiplets used within the CMF framework.

\begin{itemize}

\item Baryon matrix
\begin{equation}
B=\begin{pmatrix}
\frac{\Sigma^{0}}{\sqrt{2}}+\frac{\Lambda}{\sqrt{6}} & \Sigma^{+} & p \\
\Sigma^{-} & \frac{-\Sigma^{0}}{\sqrt{2}}+\frac{\Lambda}{\sqrt{6}} & n \\
\Xi^{-} & \Xi^{0} & -2\frac{\Lambda}{\sqrt{6}}
\end{pmatrix}\,.
\label{eq:B_matrix}
\end{equation}

\item Scalar-meson matrix
\begin{equation}
X=\begin{pmatrix}
\frac{\delta^{0}+\sigma}{\sqrt{2}} & \delta^{+} & \kappa^{+} \\
\delta^{-} & \frac{-\delta^{0}+\sigma}{\sqrt{2}} & \kappa^{0} \\
\kappa^{-} & \bar{\kappa}^{0} & \zeta
\end{pmatrix}\,.
\label{eq:X_matrix}
\end{equation}

\item Pseudoscalar-meson octet matrix
\begin{align}
&\hspace{1cm}P=\nonumber\\ 
&\begin{pmatrix}
\frac{1}{\sqrt{2}}\left(\pi^{0}+\frac{\eta^{8}}{\sqrt{1+2w^{2}}}\right) & \pi^{+} & 2\frac{K^{+}}{w+1} \\
\pi^{-} & \frac{1}{\sqrt{2}}\left(-\pi^{0}+\frac{\eta^{8}}{\sqrt{1+2w^{2}}}\right) & 2\frac{K^{0}}{w+1} \\
2\frac{K^{-}}{w+1} & 2\frac{\bar{K^{0}}}{w+1} & -\sqrt{\frac{2}{1+2w^{2}}}\eta^{8}
\end{pmatrix},\nonumber\\
\label{eq:PS_matrix}
\end{align}
where $w=\sqrt{2}\zeta_{0}/\sigma_{0}$. 

\item Pseudoscalar-meson singlet matrix
\begin{equation}
Y=\sqrt{\frac{1}{3}}\eta^{0}\begin{pmatrix}
1 & 0 & 0 \\
0 & 1 & 0 \\
0 & 0 & 1
\end{pmatrix}\,.
\label{eq:Y_matrix}
\end{equation}

\item Vector-meson matrix
\begin{equation}
V_\mu=\begin{pmatrix}
\frac{\rho_\mu^{0}+{\omega}_\mu}{\sqrt{2}} & \rho_\mu^{+} & K_\mu^{*+} \\
\rho_\mu^{-} & \frac{-\rho_\mu^{0}+{\omega}_\mu}{\sqrt{2}} & K_\mu^{*0} \\
K_\mu^{*-} & \bar{K}_\mu^{*0} & {\phi}_\mu
\end{pmatrix}\,.
\label{eq:V_matrix}
\end{equation}

\end{itemize}

\section{ Derivatives of thermal meson masses with respect to mean-fields}
\label{sec:derivative_mass_fields}
In this section, we show the derivatives of the in-medium masses of pseudoscalar (\Cref{eq:non_deg_pi_mass,eq:non_deg_Kpm_mass,eq:non_deg_K0b_mass,eq:no_mix_eta8_mass,eq:eta0_mass}) and vector (\Cref{eq:omega_mass,eq:rho_mass,eq:phi_mass,eq:Kstar_mass}) mesons computed with respect to the meson mean fields.

\subsection{Pseudoscalar mesons}

\begin{itemize}

\item  $\pi^0/\pi^+/\pi^-$:
\begin{align}
    \frac{\partial m^*_{\pi^0/\pi^+/\pi^-}}{\partial \sigma} &=\frac{1}{2 m^*_{\pi^0/\pi^+/\pi^-}} \bigg(\frac{m^2_{\pi}}{\sigma_0}\bigg)\,,
\nonumber\\
    \frac{\partial m^*_{\pi^0/\pi^+/\pi^-}}{\partial \zeta} &=  \frac{\partial m^*_{\pi^0/\pi^+/\pi^-}}{\partial \delta} =0\,,
\nonumber\\
    \frac{\partial m_{\pi^0/\pi^+/\pi^-}^{*}}{\partial {\omega}}&=\frac{\partial m_{\pi^0/\pi^+/\pi^-}^{*}}{\partial \rho}= \frac{\partial m_{\pi^0/\pi^+/\pi^-}^{*}}{\partial {\phi}}=0\,.
\end{align}
\item  $K^+/K^-$:
\begin{align}
    \frac{\partial m^*_{K^+/K^-}}{\partial \sigma} &=\frac{1}{2 m^*_{K^+/K^-}} \bigg(\frac{0.5 m^2_{K}  \sqrt{2} \left(\sqrt{2} \sigma_{0} + 2 \zeta_{0}\right)}{\left(\sigma_{0} + \sqrt{2} \zeta_{0}\right)^{2}}\bigg)\,,
\nonumber\\
    \frac{\partial m^*_{K^+/K^-}}{\partial \zeta} &= \frac{1}{2 m^*_{K^+/K^-}} \bigg(\frac{ m^2_{K}   \left(\sqrt{2} \sigma_{0} + 2 \zeta_{0}\right)}{\left(\sigma_{0} + \sqrt{2} \zeta_{0}\right)^{2}}\bigg)\,,
\nonumber\\
    \frac{\partial m^*_{K^+/K^-}}{\partial \delta} &=\frac{1}{2 m^*_{K^+/K^-}} \bigg(\frac{0.5 m^2_{K}  \sqrt{2} \left(\sqrt{2} \sigma_{0} + 2 \zeta_{0}\right)}{\left(\sigma_{0} + \sqrt{2} \zeta_{0}\right)^{2}}\bigg)\,,
\nonumber\\
    \frac{\partial m_{K^+/K^-}^{*}}{\partial {\omega}}&=\frac{\partial m_{K^+/K^-}^{*}}{\partial \rho}= \frac{\partial m_{K^+/K^-}^{*}}{\partial {\phi}}=0\,.
\end{align}
\item  $K^0/\bar K^0$:
\begin{align}
    \frac{\partial m^*_{K^0/\bar K^0}}{\partial \sigma} &=\frac{1}{2 m^*_{K^0/\bar K^0}} \bigg(\frac{0.5 m^2_{K}  \sqrt{2} \left(\sqrt{2} \sigma_{0} + 2 \zeta_{0}\right)}{\left(\sigma_{0} + \sqrt{2} \zeta_{0}\right)^{2}}\bigg)\,,
\nonumber\\
    \frac{\partial m^*_{K^0/\bar K^0}}{\partial \zeta} &= \frac{1}{2 m^*_{K^0/\bar K^0}} \bigg(\frac{ m^2_{K}   \left(\sqrt{2} \sigma_{0} + 2 \zeta_{0}\right)}{\left(\sigma_{0} + \sqrt{2} \zeta_{0}\right)^{2}}\bigg)\,,
\nonumber\\
    \frac{\partial m^*_{K^0/\bar K^0}}{\partial \delta} &=-\frac{1}{2 m^*_{K^0/\bar K^0}} \bigg(\frac{0.5 m^2_{K}  \sqrt{2} \left(\sqrt{2} \sigma_{0} + 2 \zeta_{0}\right)}{\left(\sigma_{0} + \sqrt{2} \zeta_{0}\right)^{2}}\bigg)\,,
\nonumber\\
    \frac{\partial m_{K^0/\bar K^0}^{*}}{\partial {\omega}}&=\frac{\partial m_{K^0/\bar K^0}^{*}}{\partial \rho}= \frac{\partial m_{K^0/\bar K^0}^{*}}{\partial {\phi}}=0\,.
\end{align}
\item  $\eta^8$:
\begin{align}
    \frac{\partial m^*_{\eta^8}}{\partial \sigma} &=\frac{1}{2 m^*_{\eta^8}} \bigg(\frac{m^2_{\pi}  \sigma_{0}}{\sigma_{0}^{2} + 4 \zeta_{0}^{2}}\bigg)\,,
\nonumber\\
    \frac{\partial m^*_{\eta^8}}{\partial \zeta} &= \frac{1}{2 m^*_{\eta^8}} \bigg(\frac{ \sqrt{2}  \left(\sqrt{2} m^2_{K} \left(\sqrt{2} \sigma_{0} + 2 \zeta_{0}\right) - 2 m^2_{\pi} \sigma_{0}\right)}{\sigma_{0}^{2} + 4 \zeta_{0}^{2}}\bigg)\,,
\nonumber\\
    \frac{\partial m^*_{\eta^8}}{\partial \delta} &=0\,,
\nonumber\\
    \frac{\partial m_{\eta^8}^{*}}{\partial {\omega}}&=\frac{\partial m_{\eta^8}^{*}}{\partial \rho}= \frac{\partial m_{\eta^8}^{*}}{\partial {\phi}}=0\,.
\end{align}
\item  $\eta^0$:
\begin{align}
    \frac{\partial m^*_{\eta^0}}{\partial \sigma}&=\frac{\partial m^*_{\eta^0}}{\partial \zeta}=\frac{\partial m^*_{\eta^0}}{\partial \delta}=0\,,
\nonumber\\
    \frac{\partial m_{\eta^8}^{*}}{\partial {\omega}}&=\frac{\partial m_{\eta^8}^{*}}{\partial \rho}= \frac{\partial m_{\eta^8}^{*}}{\partial {\phi}}=0\,.
\end{align}
\end{itemize}

\subsection{Vector mesons}

\begin{itemize}
\item  $\omega$:
\begin{align}
    \frac{\partial m_{{\omega}}^{*}}{\partial \sigma}&=\frac{\partial m_{{\omega}}^{*}}{\partial \zeta}=\frac{\partial m_{{\omega}}^{*}}{\partial \delta}=0\,,
\nonumber\\
    \frac{\partial m_{{\omega}}^{*}}{\partial {\omega}}&= \frac{\partial m_{{\omega}}^{*}}{\partial \rho}=0\,,
\nonumber\\
    \frac{\partial m_{{\omega}}^{*}}{\partial {\phi}}&=\frac{1}{2 m_{{\omega}}^{*}} 12g_4 \bigg(\frac{Z_{\phi} }{Z_{\omega}}\bigg){\phi} \,.
\end{align}
\item  $\phi$:
\begin{align}
    \frac{\partial m_{{\phi}}^{*}}{\partial \sigma}&=\frac{\partial m_{{\phi}}^{*}}{\partial \zeta}= \frac{\partial m_{{\phi}}^{*}}{\partial \delta}=0\,,
\nonumber\\
    \frac{\partial m_{{\phi}}^{*}}{\partial {\omega}}&=\frac{1}{2 m_{{\phi}}^{*}} 12 g_4 \bigg(\frac{Z_{\phi} }{Z_{\omega}}\bigg) {\omega}\,,
\nonumber\\
    \frac{\partial m_{{\phi}}^{*}}{\partial \rho} &=   \frac{\partial m_{{\phi}}^{*}}{\partial {\phi}}=0\,.
\end{align}
\item  $\rho$:
\begin{align}
    \frac{\partial m_{\rho}^{*}}{\partial \sigma}&=\frac{\partial m_{\rho}^{*}}{\partial \zeta}=\frac{\partial m_{\rho}^{*}}{\partial \delta}=0\,,
\nonumber\\
    \frac{\partial m_{\rho}^{*}}{\partial {\omega}}&=\frac{\partial m_{\rho}^{*}}{\partial \rho}= \frac{\partial m_{\rho}^{*}}{\partial {\phi}}=0\,.
\end{align}
\item  $K^*$:
\begin{align}
    \frac{\partial m_{K^*}^{*}}{\partial \sigma}&=\frac{\partial m_{K^*}^{*}}{\partial \zeta}=\frac{\partial m_{K^*}^{*}}{\partial \delta}=0\,,
\nonumber\\
    \frac{\partial m_{K^*}^{*}}{\partial {\omega}}&=\frac{\partial m_{K^*}^{*}}{\partial \rho}=\frac{\partial m_{K^*}^{*}}{\partial {\phi}}=0\,.
\end{align}
\end{itemize}

\bibliography{inspire}

\end{document}